\documentclass[useAMS,usenatbib,referee]{mn2e}

\title[K {\sevensize \rm I} 7699 \AA~ and related shell lines during the recent eclipse of $\epsilon$ Aurigae]{A study of K I 7699 \AA~ and related shell lines during the recent eclipse of $\epsilon$ Aurigae} 
\author[C. Muthumariappan et al.]{C. Muthumariappan$^{1}$\thanks{E-mail:
muthu@iiap.res.in}, M. Parthasarathy$^{2}$, R. Leadbeater$^{3}$, I.S. Potravnov$^{4}$ \newauthor M. Appakutty$^{5}$,  K. Jayakumar$^{5}$
\thanks{This file has been amended to highlight the proper use of \LaTeXe\ code with the class file.}\\
$^{1}$Indian Institute of Astrophysics, Bangalore 560 034, India\\
$^{2}$Inter-University Centre for Astronomy $\&$ Astrophysics (IUCAA), Post Bag 4, Ganeshkhind, Pune 411 007, India\\
$^{3}$Three Hills Observatory, The Birches,   CA7 1JF, UK \\
$^{4}$ Pulkovo Astronomical Observatory, Russian Academy of Sciences, 196140, Pulkovo, St. Petersburg, Russia \\
$^{5}$Vainu Bappu Observatory, Indian Institute of Astrophysics, Kavalur 635 701, India }

\begin{document} 

\date{Accepted ------. Received original form ------}

\pagerange{\pageref {firstpage}--\pageref{lastpage}} \pubyear{2014}

\maketitle

\label{firstpage}
 
\begin{abstract}
We report high-resolution (R = 30\ 000, 45\ 000 and 75\ 000) echelle and medium-resolution (R = 22\ 000 and 10\ 000) spectroscopic observations of the long-period, eclipsing binary $\epsilon$ Aurigae during the 2009 - 2011 eclipse. Low-excitation shell lines, viz, the K {\sevensize \rm I} line at 7699 \AA\ (with 346 data points), Cr {\sevensize \rm I} lines at 5345.807 \AA\ and 5348.326  \AA\ and Fe {\sevensize \rm I} line at 5110.435 \AA\ which originated from the disk shaped secondary, H$\alpha$ and the shell components of the Na D$_{1}$ and D$_{2}$ lines show significant variation in their shapes and radial velocities during the eclipse.  The equivalent width curve shown by the K {\sevensize \rm I} line around the ingress and egress phases indicates that the  gas density in the trailing edge is about a factor of two higher than the density in the leading edge. Using a geometrical model, in which a  homogeneous, cylindrical Keplarian disk eclipses the F0Ia primary star and the shell absorption lines originate from the gaseous atmosphere around an opaque disk, we fit the equivalent width and the radial velocity curves of the K {\sevensize \rm I} line covering the full eclipse.  A reasonably good fit can be achieved by a low-mass binary model where the mass of the central star of the disk is 5.4 M$_{\sun}$ and the mass of the primary is 2.5 M$_{\sun}$ and a disk size of 8.9 AU. The low-mass of the primary, with enhanced $s$-process elements found by Sadakane et al. (2010), supports that it is a post-AGB F supergiant. For the high-mass binary model, the modelled radial velocity curve deviates significantly from the observations. 

\end{abstract}

\section{Introduction}

$\epsilon$ Aurigae (7 Aur, HD 31964; HR 1605, $\epsilon$ Aur hereafter) is an eclipsing binary with the longest known orbital period of 27.1 years and showing an eclipse depth of 0.75 mag in the optical and almost a year long duration of totality phase. The eclipse depth is independent of wavelength from $\sim$ 1400 \AA\ to below the far-IR wavelength, but the depth, the duration of the eclipse and comparable masses of the components imply that the components should be almost equally bright. However, no secondary eclipse was observed.  Eclipse characteristics indicate that the occulting object is very elongated parallel to the binary orbit and it must be huge to block half of the primary's light. It was suggested that the primary is occulted by a disk shaped secondary causing the two year long eclipse (Huang 1965, Kopal 1971, Gyldenkarne 1970, Wright 1970, Sahade \& Wood 1978). IR observations long-ward of 5$\mu$ region made by Backman et al. (1984) revealed an excess emission in the infrared which can be attributed to the thermal radiation from the dust in the disk with a temperature of $\sim$ 550 K (also see Hoard et al. 2010; Muthumariappan \& Parthasarathy 2012). 

The K {\sevensize \rm I} 7699 \AA\ line variation and hence the presence of neutral gas in and around the disk shaped secondary of $\epsilon$ Aur was for the first time discovered by Parthasarathy (1982) (see also Stencel 1982, Parthasarathy \& Lambert (1983a,b,c) from the systematic increase in the strength of K {\sevensize \rm I} 7699 \AA~ line during the 1982 - 1984 eclipse. Hack \& Selvelli (1979) first observed UV excess in the $IUE$ UV spectrum of $\epsilon$ Aur. The  $IUE$ UV spectra obtained during the 1982 - 1984 eclipse also show UV excess and indicates a B5 V star at the centre of the disk (Parthasarathy \& Lambert 1983d; Boehm, Ferluga \& Hack 1985; Chapman, Kondo \& Stencel 1983; Altner et al. 1986, Ake 1984, 2006). Light and colour variations during the eclipse of $\epsilon$ Aur was discussed by Parthasarathy \& Frueh (1986). $\epsilon$ Aur has recently finished an eclipse cycle which started in 2009 August and ended in 2011 August. Numerous photometric, spectroscopic and interferometric observations were carried out from UV to IR during 1982 - 1984 and 2009 - 2011 eclipses; Stencel (2012) presented a comprehensive summary of the recent eclipse results (see Table. 1 for photometric contact details). The interferometric images published by Kloppenborg et al. (2010) show that the eclipsing object is clearly disk shaped. Kloppenborg et al. (2012) presented the light-curve variation of $\epsilon$ Aur during the 2009 - 2011 eclipse. 

Recently, Stefanik et al. (2010) reported an updated single-lined spectroscopic binary solution for the orbit of the F0Ia primary based on 20 years of monitoring at the CfA, combined with historical velocity observations dating back to 1897. They presented two solutions. One uses the velocities outside the eclipse phases together
with mid-times of previous eclipses, from photometry dating back to 1842, which provide strongest constraint on the ephemeris. From this they find an orbital period of 9882 days (27.0938 years), an orbital eccentricity of 0.29, system radial velocity of --2.26 $\pm$ 0.15 km s$^{-1}$ and a mass function of 2.51 M$_{\sun}$ for the binary using a combined fit to the data (where the eclipse times were incorporated with the radial velocities into the least-square fit). By using only radial velocities they find that the predicted middle of the current eclipse is nine months earlier than estimated from photometric study, implying that the gravitating companion in the binary is not the same as the eclipsing object. They conclude that, the purely spectroscopic solution may be biased by perturbations in the velocities due to the short-period oscillations of the F0Ia primary star. Other notable recent results are the interferometric studies by Stencel et al. (2008), photospheric abundance analysis of Sadakane et al. (2010), spectral and photometric analysis presented by Chadima et al. (2011) and the detailed infrared studies of $\epsilon$ Aur during the 2009 - 2011 eclipse (Stencel et al. 2011).  More recently, Griffin \& Stencel (2013) conclude that the structure of the disk does not alter appreciably on a time scale of a century and they discovered a mass transfer stream from the F0Ia star on to the disk. They suggested that the F0Ia may be a horizontal branch star. The exact nature of the disk and the masses and the evolutionary nature of the binary components are not yet known well.

Orbital characteristics and spectral properties of the primary indicate two different models for the $\epsilon$ Aur system. One is the high-mass star model with the F0Ia primary having a mass of 15M$_{\sun}$ (Carroll et al. 1990) in which the disk is a  proto-planetary or proto-stellar disk and it is most unlikely to have resulted from the primary overflowing its Roche lobe and transferring mass to the secondary.  The other one is the low-mass star model (Eggleton \& Pringle 1984) in which the F0Ia primary has a mass of 4M$_{\sun}$ or less, which supports an accretion disk for $\epsilon$ Aur which is a remnant of post-main sequence mass transfer from the primary to the secondary. Distance measurement made by Hipparcos (650 pc, Perryman et al. 1997) and in turn an estimate of intrinsic luminosity of the primary does not favour a high-mass star model. However, Hipparcos distance estimate has significant error and it was argued based on the interstellar absorption and reddening that the distance to $\epsilon$ Aur is 1500 pc (Guinan et al. 2012), which will set a large intrinsic luminosity for the system and indicates a high-mass star system. Though this issue is not yet resolved, it is now clear that $\epsilon$ Aur is a  binary system for which the brighter F0Ia component can be less massive than its companion (Parthasarathy \& Pottasch 1986).  

Disks are among most common astrophysical systems. The $\epsilon$ Aur disk has been described variously as thick (Huang 1965) or thin, flat or twisted, opaque or semi transparent, fully solid or possessing a central hole. Hoard et al. (2012) showed that  their photometric data obtained with the $Herschel$ $Space$ $Observatory$ are consistent with a cool disk model. Similar results were found from their 2-D radiative transfer modelling of the dusty disk by Muthumariappan \& Parthasarathy (2012). They also discussed that the disk is not continuous, but has a void of radius 2 AU at its centre. If the low-mass star model suits for $\epsilon$ Aur system, then it offers an unique opportunity to study in detail a new class of disks which are associated with post-AGB stars.

$\epsilon$ Aur shows a shell spectrum during the eclipse which manifests itself by the presence of additional absorption lines arising from the optically thin outer atmosphere of the disk. Numerous absorption lines are seen in the optical spectrum of $\epsilon$ Aur which showed significant variation during the recent eclipse (Leadbeater et al. 2012). The vast majority of the primary's lines are accompanied by one or two velocity shifted absorption components formed as the light of F0Ia star passes through the neutral gas in and around the disk. These spectral lines which show absorption due to the disk and display spectral variation during the eclipse can be very useful tools to trace the disk structure and to derive its parameters. 

The low-excitation K {\sevensize \rm I} line at 7699 \AA\ is one of the key spectral lines to understand the disk as, unlike other lines mentioned above, this line was absent (except the interstellar component) in the out-of-eclipse spectrum of the F0Ia star and shows systematic variation in strength during the eclipse phase (Parthasarathy 1982, Parthasarathy \& Lambert  1983a,b,c, Lambert \& Sawyer 1986, Leadbeater et al. 2012). This line has been used to study the disk structures (Leadbeater \& Stencel 2010; Potravnov \& Grinin 2013). Gas associated with the secondary is most clearly traced by the K {\sevensize \rm I} resonance lines at 7664 and 7699 \AA\ originated from in and around the disk shaped secondary. The stronger line at 7664 \AA\ is generally blended with strong telluric lines of O$_{2}$. It was also discussed that several low-excitation lines of Fe peak elements like Cr {\sevensize \rm I}, Sc {\sevensize \rm I}, Ti {\sevensize \rm I} are most likely shell lines which can be used to trace the disk structure (Sadakane et al. 2013). However, unlike the  K {\sevensize \rm I} at 7699 \AA\ , these shell lines are not present in the spectra throughout the eclipse. Fe I lines appear from the mid-eclipse to the third contact and Cr {\sevensize \rm I} lines are present from third contact to fourth contact. All other lines appear only for a very short duration near the third contact (see also Fig. 4 of Sadakane et al. 2013).

In this paper we study the variations shown by the shell lines K {\sevensize \rm I} at 7699 \AA\ (excitation potential 1.6 eV; K {\sevensize \rm I} line hereafter), Fe {\sevensize \rm I} at 5110.435 \AA\  and 5434.53 \AA\  (excitation potential 1.01 eV; Fe {\sevensize \rm I} lines hereafter) and Cr {\sevensize \rm I} at 5345.807 \AA\ and 5348.326 \AA\ (excitation potential 1.00 eV; Cr {\sevensize \rm I} lines hereafter) in their strength, line profile shape and in their radial velocity using high- and medium-resolution spectroscopic observations obtained during the 2009 - 2011 eclipse. We also present the variation shown by the Na D and H$\alpha$ lines.  Using these results we construct a model for the eclipse of $\epsilon$ Aur and derive the geometrical parameters of the disk. We further constrain the masses of the stellar components using our model and discuss the evolutionary status of $\epsilon$ Aur and the origin of its disk. 

\section {Observations and data reduction} 

The high-resolution spectroscopic observations used in this study were obtained from a fibre fed echelle spectrograph, attached to the prime focus of the Vainu Bappu Telescope (VBT) located at the Vainu Bappu Observatory, Kavalur, India. The echelle spectra in the 4000 \AA\ to 9000 \AA\ wavelength region were recorded on a 4k $\times$ 4k CCD detector with a pixel size of 12 $\mu$m. These observations were taken in stable and clear sky conditions with exposure times varying from 3 to 15 minutes for each spectrum. We have recorded 52 echelle spectra during December 2010 to April 2012. Our spectra hence were taken before the third contact (27 February 2011) to the end of the eclipse (26 August 2011). In addition, some spectra were recorded during eight months period after the end of the eclipse. Most of the spectra (33 of 52) were taken with the slit mode giving a spectral resolution  of $\sim$ 75\ 000. Some of the spectra (19 of 52) were  obtained in the slit-less mode which gives a resolution of 30\ 000.  75\ 000  is the highest spectral resolution ever obtained on this object during its 2009 - 2011 eclipse. 

These data were reduced and analysed with a pipelined data analysis steps using $IRAF$ software. We have taken the following data reduction sequence: bias subtraction, flat fielding, scattered light subtraction, spectral extraction and wavelength calibration.  The wavelength calibration was done using Th-Ar comparison spectra which were taken along with $\epsilon$ Aur observations. The observed velocities are then converted to the helio-centric values. The telluric lines present at different spectral locations were used to find the error in our radial velocity calibration. The maximum error in radial velocities in the slit mode derived from un-blended lines is found to be 78 m sec$^{-1}$ and for the blended lines we have employed multiple gaussian fitting which leads to a maximum error of 100 m sec$^{-1}$ in this observing mode. Spectra taken in the slit-less mode has a maximum error of 320 m sec$^{-1}$ for the un-blended lines and for the blended lines the maximum error is 400 m sec$^{-1}$.  The signal-to-noise ratio of our observations varies from 200 to 400. Table 2 lists the measurements of K {\sevensize \rm I} line equivalent widths and radial velocities obtained from VBT data.

We have combined these observations with 247 measurements of the K {\sevensize \rm I} line equivalent widths and radial velocities made by Leadbeater at Three Hills Observatory (THO) using a LHIRES III  spectrograph (with a spectral resolution of $R$ = 22 000) with 200 mm and 280 mm telescopes. The measurements cover the full eclipse period from 29 May 2009 to 21 March 2012 which are listed in Table 3.  In this line, a constant interstellar component is blended with the system component for most of the eclipse. This was removed before measuring the line parameters (based on measurements made outside the eclipse). Further details of the data reduction can be found in Leadbeater et al. (2012). In addition, we have included the values of equivalent widths and radial velocities of K {\sevensize \rm I} lines measured from 47 high-resolution ($R$ = 45 000) spectra (after removing the interstellar component) obtained  by Potravnov \& Grinin (2013) using MAESTRO echelle spectrograph attached to the 2 m telescope at the Terskol Observatory (TO). Table 4 lists the measurements of K {\sevensize \rm I} line equivalent widths and radial velocities obtained from TO data. Our data includes a total of 346 values of equivalent widths and of radial velocities of the K {\sevensize \rm I} 7699 \AA\ line covering entire eclipse during 2009 to 2011. Table 5. shows the equivalent width and radial velocity values for the Fe {\sevensize \rm I}  5110 \AA\ line extracted from spectra submitted as part of the International $\epsilon$ Aurigae Campaign 2009 - 2011 and available on-line at ${www.threehillsobservatory.co.uk/epsaur_{} spectra.htm}$. The spectra were recorded using eShel echelle spectrographs at a resolving power $R$ = 10 000 (${www.shelyak.com}$). Telescope aperture was 280 mm (Buil, Thizy) and 355 mm (Garrel). For the Na D$_{2}$ line, the equivalent width and radial velocity values from Gorodenski (2012; observations made with a spectral resolution of $R$ = 18 000) were included in our study.

\section {Results} 

\subsection{Spectral line variations}

All the spectral lines presented in this study from three different observatories  show significant variation in line profile shape and in radial velocity during the eclipse. A time series of a few selected spectra taken near K {\sevensize \rm I},  Fe {\sevensize \rm I} 5110 \AA\ , Fe {\sevensize \rm I} 5440 \AA\, Cr {\sevensize \rm I},  H$\alpha$ and Na D lines are shown respectively in Fig. 1, Fig. 2, Fig. 3, Fig. 4. Fig. 5, and in Fig. 6. In addition, also shown in the figures are the spectra taken at these lines after the end of the eclipse (on 7 April 2012) for comparison. All the spectra shown in the figures were taken using VBT echelle spectrograph with a spectral resolution of 75\ 000. Interstellar absorption components were well resolved and hence were not removed from these spectra.

\subsubsection{K {\sevensize \rm I} line and other shell lines}

The shell lines were seen in the spectrum only during or near the eclipse. The K {\sevensize \rm I} line was present throughout the eclipse and was seen even after the eclipse for a few months. This line shows a blue asymmetry until 18 February 2011, and this asymmetry breaks into two blended components thereafter. The line profiles are increasingly blue shifted until 18 March 2011 (near the third contact), then onwards the trend becomes opposite (see Fig. 1). As shown in Fig. 2, and Fig. 3, the Fe {\sevensize \rm I} lines are blue shifted until 24 March 2011 and shows an opposite trend thereafter. The absorption strength of the Fe {\sevensize \rm I} lines also increases towards this turning point and the line strength is maximum on 2 April 2011, nearly a month after the third contact.  However, after this date the Fe {\sevensize \rm I} line is very weak. Cr {\sevensize \rm I} lines show very similar behaviour with the Fe {\sevensize \rm I} lines both in radial velocity pattern and in absorption strength variation (see Fig. 4).  The above findings on the Fe {\sevensize \rm I} and Cr {\sevensize \rm I} lines are in good agreement with the results of Sadakane et al. (2013).  Unlike the K {\sevensize \rm I} line, the other three shell lines do not show any noticeable line profile asymmetry. These lines were seen in the spectra only for a short duration near the third contact in our data and also in the archive. This was also noted by Sadakane et al. (2013). 

\subsubsection {H$\alpha$ and Na D lines}

The H$\alpha$ absorption line profiles are broad and display several sub-structures during the eclipse (see Fig. 5). Spectra which were obtained after the eclipse show relatively narrower line profiles with no noticeable sub-structures. H$\alpha$ line profiles are mostly asymmetrical; the violet wing of the profile shows a steep fall and the red wing gradually increases to the continuum level. The variation shown by these profiles is quite complex during the eclipse. The H$\alpha$ absorption line profile consists of both photospheric and shell components. Weak emission components are often seen at the violet and at the red edges of the absorption profile. The red edge emission component is relatively brighter than the violet edge component. The H$\alpha$ absorption component also shows variability outside the eclipse phase (Guangwei et al. 1994). The interpretation of H$\alpha$ line profile variation during the eclipse hence needs to take the variations outside the eclipse into account. For example, the complex behaviour of the H$\alpha$ line during the eclipse was examined by Chadima et al. (2011) after subtracting the mean out-of-eclipse profile. As it can be seen from the spectra shown in Fig 6, the Na doublet lines are quite broad (and some are possibly saturated) during the eclipse and do not have sharp dip. Unlike the H$\alpha$ line, these absorption lines have a more symmetrical structure and do not show any sub-structures. They also do not have emission components at the edges of the absorption lines.
 
\subsection{Analysis}

 To understand the disk structure, we have analysed the variations shown by the shell absorption lines in their radial velocities (RVs), equivalent widths (EWs) and in full-width at half-maxima (FWHMs) during the eclipse. The values of these parameters were measured after fitting a continuum to the spectrum using a polynomial regression model. The beginning and the end points of a line are defined where the line is judged to come in contact with, or cross through the estimated continuum. The split seen in the K {\sevensize \rm I} line near the third contact was not taken into account for our analysis and we have considered this as a single line with mean velocity of its components. The wavelength, EW and FWHM of a symmetrical line profile were obtained by making a gaussian fit. Wherever an asymmetry is seen in the line profile we have employed multi-gaussian fitting method and the mean position of the velocity components was taken to estimate the RV. The EW of individual components were quadratically added to get the total value. The width of a line profile at its half maximum is taken as the FWHM, for both symmetrical the asymmetrical profiles. In the following sections we discuss the variations shown by these parameters for different spectral lines during the eclipse. The K {\sevensize \rm I} line is strong enough throughout the eclipse so that we could measure these parameters in all the spectra. However, the other shell lines show absorption features with measurable strengths only during a part of the eclipse phase (near the third contact).  The values of 
EW and RV values for the K {\sevensize \rm I} line listed in Table 4, are re-measured from the spectra obtained by Potravnov \& Grinin (2013) after the subtraction of the interstellar component.

\subsubsection{Equivalent Width variation}

The EW at a particular phase of eclipse measures the quantity of absorbing matter in the disk along the line-of-sight to the F0Ia star. The variation of EW against the Julian date during the eclipse for three shell lines and Na D$_{2}$ line considered in this study are shown in Fig. 7.  The interstellar absorption component near the 
K {\sevensize \rm I} line is well resolved in all the VBT spectra, hence the EW of the disk component can be measured without any confusion. The interstellar component of the K {\sevensize \rm I} line has a mean EW of 0.1036 $\pm$ 0.0007 \AA\ , as measured from the VBT spectra.  The EW of the interstellar K {\sevensize \rm I} line measured from the THO observations is 0.111 \AA\ and Potravnov \& Grinin (2013) reported a value of 0.119 \AA\ . The estimated uncertainty in EW measurements from VBT spectra,  as arrived from repeated measurements, is less than 4$\%$ for the K {\sevensize \rm I} and Na D lines. The error could be larger (upto 7$\%$) for the Fe {\sevensize \rm I}, Cr {\sevensize \rm I} lines as their absorption strengths are weak.

The K {\sevensize \rm I} line is nearly, but not quite optically thin during the eclipse. The EW curve of the K {\sevensize \rm I} line shows a dip before 25 December 2010 (see Table 2), which was not covered by VBT observations.  The VBT data do not cover the second contact when the EW reached its first maximum and the mid-eclipse phase where it reached a dip. The second contact as seen from the K {\sevensize \rm I} line observations of THO was on JD 2455250 and the EW on this date is 0.476 \AA\ . The closest point to the second contact in the TO data is on JD 2455275 on which the measured EW is 0.4 \AA\ .  The mid-eclipse seen in the   K {\sevensize \rm I} line had occurred on 27 July 2010 (JD 2455405) as it is seen from the spectroscopic data of the THO (see Table 3) when the EW has a value of 0.332 \AA\ . The EW steeply rises to 0.860 \AA\ on 20 Feb 2011 (JD 2455613), which is the nearest point we have to the third contact in the VBT data (see Table 2). The third contact, as seen from the spectroscopic data of THO, was on 26 Feb 2010 (JD 2455619) on which the measured EW was 0.796 \AA\ (see Table 3). This disk-origin EW decreases gradually then after to a lowest value of 0.100 \AA\  on 14 December 2011 (JD 2455909) in the VBT data, and a value of 0.114 \AA\ on JD 2455888 in the TO data. In the THO data, the lowest EW value originated from the disk is 0.021 \AA\ which was obtained on JD 2456005. The EW curve of Na D$_{2}$ line steeply rises to 1.83 \AA\ on 29 March 2011 (JD 2455650) in the VBT data then it decreases gradually to its lowest value of 1.15 \AA\ which occurred on 5 December 2011 (JD 2455900).

The end of the optical photometric eclipse is on 26 August 2011 (Stencel 2012) and the K {\sevensize \rm I} line in principle should disappear completely, except for the interstellar part, after this date. However, the decline in EW continues well after the fourth contact, for about three months after the end of the eclipse. The spectral line is seen in THO data with measurable strength until JD 2456005, when the primary F0Ia star should be far clear from the disk. 

The Fe {\sevensize \rm I} line and the Cr {\sevensize \rm I} lines also seem to show a similar trend within the eclipse phase where they were observed. The maximum EW values for the Fe {\sevensize \rm I} line and Cr {\sevensize \rm I} lines obtained from VBT data were near to the third contact (JD 2455612) and their respective values are 0.071 \AA\  and 0.052 \AA\ . This is very close to the date on which K {\sevensize \rm I} line also showed its maximum EW.  As noted earlier, these low-excitation lines are seen only for a short duration near the third contact and so there are no measurements available during the earlier eclipse phases to compare them with the K {\sevensize \rm I} line EW curve. The EW curves of all shell lines do not show a smooth pattern and they display jumps at various eclipse phases. This indicates a non-uniform neutral gas density distribution and presence of sub-structures in and around the disk (see also Leadbeater et al. 2012). However VBT observations show that the positions of these jumps in EW curves are not same for K {\sevensize \rm I}, Fe {\sevensize \rm I}, Cr {\sevensize \rm I} and Na D$_{2}$ lines. 

\subsubsection{Radial Velocity variation}

The RV at a particular phase of the eclipse measures the mean velocity of the absorbing material of the disk in front of the F0Ia star, projected along the line-of-sight. This includes the orbital motion of the disk around the  centre of mass of the binary system and the rotation motion of the disk around the unseen secondary star. The radial velocity curve can be used to deduce the rotation curve of the disk, as near the mid-eclipse the line-of-sight projected orbital velocity will be negligible but the observed RV of the disk is still larger than the RV of the F0Ia star. The variation of RV against the Julian date during the eclipse for the three shell lines and Na D$_{2}$ line considered in this study are shown in Fig. 8.  

The RV curves of K {\sevensize \rm I} and Na D$_{2}$ lines show similar trend: both show their lowest values around JD 2455613 (near the third contact) in VBT data and it increases there after. However, the lowest RV seen in these lines are different: --41.42 km $s^{-1}$ for K {\sevensize \rm I} line (see Table 2) and --24.12 km s$^{-1}$ for Na D$_{2}$ line. The lowest RV obtained from the THO data at K {\sevensize \rm I} line was on JD 2455649 which is --40.5, and on the third contact (JD 2545619) its value is --38.6 (see Table 3). As the VBT data do not cover the second contact, the maximum value of RV could not be obtained. The second contact as seen from the K {\sevensize \rm I} line observations of THO was on JD 2455250 with a RV of 17.5 km s$^{-1}$. However, the largest RV measured from K {\sevensize \rm I} line from THO data is 20.3 km s$^{-1}$ which was on JD 2455205. The largest RV measured from the Na D$_{2}$ line is 16.53 km s$^{-1}$ (Gorodenski 2012). The data obtained from TO at K {\sevensize \rm I} line do not have a RV point close to the third contact but the RV value obtained close to the second contact is 21.8 km s$^{-1}$ which occurred on JD 2455275. Radial velocities measured from the Fe {\sevensize \rm I} line and Cr {\sevensize \rm I} line do not show their minimum values on JD 2455613. 
The minimum RV observed for  the Fe {\sevensize \rm I} line is --45.5 km s$^{-1}$ which occurred on JD 2455649 (near the third contact). The minimum RV measured for the Cr {\sevensize \rm  I} line was on the same date. Its value is --46.8 km s$^{-1}$, which is a little larger than the value shown by the Fe {\sevensize \rm I} line. Both the Fe {\sevensize \rm I} and Cr {\sevensize \rm I} lines show minimum velocities which are lower than that observed for the K {\sevensize \rm I} line, however,  the minimum velocities of these two shell lines were seen 30 days after the occurrence of the minimum velocity of the K {\sevensize \rm I} line. These findings are in good agreement with Sadakane et al. (2013). The RV curve derived from the K {\sevensize \rm I} line (Fig. 8) shows maximum and minimum values which are not symmetric with respect to the radial velocity of the binary system (--2.26 km s$^{-1}$) as also can be seen in Fig. 6 of Sadakane et al. (2010). This radial velocity asymmetry is also seen in the RV curve of the Na D$_{2}$ line (Fig. 11), however, not as large as it is seen for the K {\sevensize \rm I} line. A possible explanation for this asymmetry is discussed in section 4.2.

\subsubsection{Full-Width at Half-Maximum variation}

The FWHM traces the broadening of the spectral lines at a particular phase of the eclipse. The variations of the FWHM against the Julian date are shown in Fig. 9 for the three shell lines and the Na D$_{2}$ line considered in our study. The FWHM measured from the K {\sevensize \rm I} line and the Na D$_{2}$ line decrease to their minimum values which had occurred near JD 2455900 (about three months after the end of optical eclipse) and they increase then after. The minimum value of FWHM seen for the K {\sevensize \rm I} line is 0.286 \AA\ and for the Na D$_{2}$ line it is 1.02 \AA\ .   K {\sevensize \rm I} line FWHM could be measured in our data until JD 2456005. This steady increase of FWHM curve may show the presence of matter radially outside the main disk, possibly in the form of a ring with large velocity dispersion. As seen earlier, evidence for this extended disk was also seen in the EW curves of the shell lines. Though a trend of decreasing FWHM was seen for the Fe {\sevensize \rm I} and Cr {\sevensize \rm I} lines, similar to the K {\sevensize \rm I} and Na D$_{2}$ lines, there are no reliable value of FWHM available beyond JD 2455660 due to the weakness of the spectral line after the third contact. The minimum FWHM for the Fe {\sevensize \rm I} line and  Cr {\sevensize \rm I} lines which were observed near the third contact have their respective values of 0.25 \AA\ (14.6 km s$^{-1}$) and 0.26 \AA\ (14.5 km s$^{-1}$).

\section{A geometrical model for the eclipse of $\epsilon$ Aur} 

\subsection{Model Description}
We attempt here to understand the trends shown by the EW and the RV curves of the shell lines of $\epsilon$ Aur during the eclipse using a geometrical model and analytically solving the line parameters in the optically thin limit. Using this model we further try to constrain the disk geometrical parameters  and the stellar masses of the binary system. In this model we assume that the disk is cylindrical in shape in which gas is distributed uniformly. We do not address here the disk sub-structures shown by the spectral lines, however, to understand the trends shown by the EW and RV curves of the shell lines this can be sufficient. The gas in the disk is in Keplerian rotation around the star located at the centre of the disk. This assumption establishes a relation between the RV due to the disk rotation, the size of the disk and the mass of the secondary. The optically visible F0Ia star and the central star of the disk rotate around their centre of mass in an orbit which has an inclination of 90$^{o}$. The inner disk is opaque and seen edge-on (as suggested by earlier studies; Hoard et al. 2010; Muthumariappan \& Parthasarathy 2012) and the shell absorption lines arise from the peripheral and transparent atmosphere of the disk located around the opaque region. We consider that the disk has an outer diameter (the diameter of the atmosphere) of $D_{out}$ and has an inner diameter (the diameter of the opaque disk) of $D_{in}$. The shell absorption lines are formed between $D_{out}$ and $D_{in}$. The disk has a central void of diameter 4 AU (Muthumariappan \& Parthasarathy 2012) which is surrounded by the opaque disk and the gaseous atmosphere surrounds the opaque disk, hence, $D_{out} \ge D_{in} \ge$ 4 AU.
We further assume here that the disk diameter is significantly larger than the photospheric size of the F0Ia star. Kloppenborg et al. (2010) showed that the angular diameter of the opaque disk is about six times the angular diameter of the star. Hence, $D_{out}$ is larger than six times the size of the photosphere and this assumption is appropriate. 

As the disk passes in front of the F0Ia star, neutral gas which is present in the disk column overlapping the line-of-sight cone absorbs the continuum flux from the F0Ia star (primary) producing the observed shell lines. The thickness of the disk is taken to be constant. The total light absorbed by an infinitesimal cell element at a position $x$ and $y$ from the central star (secondary) of the disk ($x$ is normal and $y$ is parallel to the line of sight), is hence


\begin{equation}
EW = K \int \rho(x,y) dxdy       \\
\end{equation}

where, $K$ is a constant and $\rho(x,y)$ is the density of the cell. Integrating this value over the line-of-sight column will give the total EW. The integration limits of $y$ are from $\sqrt{D_{in}^{2}-x^{2}}$ to $\sqrt{D_{out}^{2}-x^{2}}$. The integration limits of $x$ depends on the line-of-sight cone size of the primary, which in turn depend on the distance. But the computed value is normalised with respect to the peak value and the distance does not constrain on the shape of EW curve. The variation of EW with the  phase of the eclipse as the disk moves in front of the F0Ia star produces the EW curve. 

The radial velocity of an infinitesimal cell element is contributed by its line-of-sight projected Keplerian velocity around the secondary ($v_{r,disk}$), the line-of-sight projected orbital velocity of the secondary around the centre of mass of the binary system ($v_{r,sec}$) and the system radial velocity ($v_{r}$). We included a constant value $v_{offset}$ to fit the RV curve (see also section 4.2). If $m_{1}$ and $m_{2}$ are respectively the masses of the primary and the secondary, the RV of a cell element at a distance $d$ from the centre of the disk which makes an angle $\delta$ with the line-of-sight is

\begin{equation}
RV = v_{r,disk} + v_{r,sec} + v_{r} + v_{offset}   \\
\end{equation}

where,

\begin{equation}
v_{r,disk} =    \sqrt{Gm_{2}/d} \times cos\delta   \\
\end{equation}

The line-of-sight projected orbital velocity of the secondary around the centre of mass can be found as follows. The disk comes along the line-of-sight to the primary during the mid-eclipse and the model requires that the secondary is in its apastron point of its orbit around the centre of mass of the binary system during the mid-eclipse (fitting EW and RV curves could not be achieved otherwise using the allowed variable parameters). The distance from the focus to the apastron of the secondary's elliptical orbit with a period $P$ and an eccentricity $e$ is given by 

\begin{equation}
a_{f} = [m_{1}(1+e)/(m_{1}+m_{2})] \times [P^{2}G(m_{1}+m_{2})/4\pi^2]^{1/3}
\end{equation}


A similar relation for the primary can be obtained by interchanging $m_{1}$ and $m_{2}$. The radial distance ($r_{sec}$) and the velocity ($v_{sec}$) of the secondary at an angular location $\theta$ (measured from $a_{f}$ to the radial vector to the centre of mass; the angular location of the primary is 180$^{o}$ + $\theta$) is 
  
\begin{equation}
 r_{sec} =  a_{f}(1-e)/(1 -ecos\theta) 
\end{equation}

\begin{equation}
 v_{sec} = \sqrt{[G(m_{1}+m_{2})/a_{f}(1-e)]\times[1+e^{2}-2ecos\theta]}
\end{equation}

The angle between this velocity vector and the line perpendicular to the radial vector $\psi$ is given as

\begin{equation}
 tan\psi = esin\theta /(1 + ecos\theta)
\end{equation}

The line-of-sight component of the secondary's orbital velocity is hence 

\begin{equation}
v_{r,sec} = v_{sec}cos\psi sin\theta
\end{equation}

The density-weighted values of $v_{r,disk}$ of the cell elements along the line-of-sight cone has a peak value, which depends on $D_{in}$, $D_{out}$ and the density function. This is added with $v_{r,sec}$, $v_{r}$ and $v_{offset}$ to get the observed radial velocity at any instant. The time variation of RV as the disk eclipses the primary gives the observed RV curve. The primary moves in the opposite direction with the same angular velocity of the secondary with respect the the centre of mass. The ingress and egress times of the eclipse take this into account. Fig. 10 shows the variation of line-of-sight integrated radial velocity components of $v_{r,disk}$ ,$v_{r,sec}$ and their addition against the eclipse phase. {\bf As it can be seen from equations 1 to 8, only the disk geometrical parameters, the radial density function and the stellar masses are used to fit the RV curve and the shape of the EW curve. If the angular diameter of the optical disk is known accurately, then the distance is given from the disk linear diameter derived from the model. Further, the masses of the stellar components depend on the distance (higher mass for  a larger distance) due to the observed luminosity of the primary and the binary mass function.  Hence, the distance is implied by the outcome of the model, though it is not an input variable.}

$D_{in}$ and $D_{out}$ constrain both the EW and RV curves. As it can be seen from equation 2., the RV curve is also determined by the secondary mass. The diameter of the disk is related to $r_{sec}$ and also to $a_{f}$ as the eclipse phases are fixed and hence the EW is indirectly connected to the primary to the total stellar mass ratio. The RV curve is strongly stellar mass dependent. If the geometrical and density parameters are constrained from the EW curve then fitting the RV curve will constrain the secondary mass. The variable parameters of our model are $D_{in}$, D$_{out}$ and the secondary mass and we worked for a model which can fit both the EW and the RV curves simultaneously. The radial density function in the disk is taken as the density function in the disk mid-plane of the accretion disk given by Lynden-Bell $\&$ Pringle (1974).   

The mass function of the binary system, its orbital period and the eccentricity of the orbit were taken from Stefanik et al. (2010) corresponding to their combined fit. These values are 2.51 M$_{\sun}$, 9896 days and 0.227 respectively. The masses of the component stars allowed by the mass function are from 5.0 to 14.3 M$_{\sun}$ for the secondary and correspondingly from 2.0 to 19.9 M$_{\sun}$ for the F0Ia primary (taking the orbital inclination as 90$^{o}$). For a set of input values of the stellar masses allowed by the mass function, the semi-major axis of the secondary's orbit was calculated using the orbital period. From this, the apastron radius  and in turn the radial velocity were estimated as described above. This was exercised for different sets of allowed stellar masses to reach a best fit to the observed RV curve using the disk inner and outer diameter derived from the EW fit.

\subsection{Model results}

Different models were computed to fit the EW and the RV curves of K {\sevensize \rm I}, Fe {\sevensize \rm I} and Na D$_{2}$ lines during the eclipse. For the Fe {\sevensize \rm I} line, significant absorption was seen from the mid-eclipse to the end of the eclipse. We could not fit these curves for the Cr {\sevensize \rm I} lines, as discussed earlier, these low-excitation lines appear only for a very short duration near the third contact and there are no sufficient data points available tracing the full eclipse phase to perform a model fit.  The best-fit model EW and RV curves are plotted against their observations in Fig. 11 and their model parameters are listed in Table 6. As it can be seen in Fig 11, while the RV curve of the K {\sevensize \rm I} line fits well by the model (by making a velocity offset, see below), its observed EW curve shows an asymmetry which could not be reproduced by the model. The ingress region of the disk deviates significantly from the model fit and the egress region makes a more reasonable fit by an uniform disk model. The EW points after JD 2455800 are quite far from the model estimation. The EW and RV curves of Na D$_{2}$ line deviate quite significantly from the model. In addition to the observed asymmetry between the egress and the ingress seen in Na D$_{2}$ EW curve which is not reproduced in the model, the model curve is narrower than the observed one in the mid-eclipse to the end of the eclipse region. 

The RV curves of the K {\sevensize \rm I} and the Na D$_{2}$ lines also show asymmetry: to fit the RV curve of the K {\sevensize \rm I} and Fe {\sevensize \rm I} lines, offset values of --4.0 km s$^{-1}$ and --3.0 km s$^{-1}$ were required to be added to their respective model curves. The RV curve of the Na D$_{2}$ line looks relatively more symmetrical, which needed an offset value of $\sim$ --1.0 km s$^{-1}$. This may indicate that the disk is not circular rather it is elliptical in shape and the central star is located in the focus located away from the observer. The major axis of the elliptical disk should also have an inclination with respect to the line-of-sight. Such a system can make a larger radial velocity towards the observer and a smaller radial velocity away from the observer when compared to a circular disk. The combined RV curve can be fitted reasonably well by just making an offset to the centre of mass velocity. The offset value is significantly smaller for the RV curve of the Na D$_{2}$ line than for the K {\sevensize \rm I} and Fe {\sevensize \rm I} lines.  Though the RV points of Na D$_{2}$ line before JD 2455300 and after JD 2455700 could not be fit by the model and the overall fit also is not as good as that for the K {\sevensize \rm I} and Fe {\sevensize \rm I} lines, the difference seen in the offset is significant. A possible explanation could be that the region of the disk traced by the Na D$_{2}$ line is relatively more circular.

The K {\sevensize \rm I} line absorption occurs in the disk from $D_{in}$ = 5.8 AU to $D_{out}$ = 8.9 AU, whereas, for the Fe {\sevensize \rm I} line the absorption occurs between $D_{in}$ = 5.4 AU and $D_{out}$ = 8.7 AU. For the Na D$_{2}$ line, the absorption occurs in a region of the disk with $D_{in}$ = 4.1 AU to $D_{out}$ $\sim$ 8.1 AU. These model parameters of the Na D$_{2}$ line RV curve should be taken with caution, as we could not reach a good fit to the observations. The best possible fit shows that the outer and inner diameters of the disk traced by the Na D$_{2}$ line is smaller than the disk diameters modelled from the other two lines. The mass of the B5V star at the centre of the disk derived from our model is 5.4 M$_{\sun}$ and the corresponding mass of the F0Ia primary, given by the mass function, is 2.5 M$_{\sun}$. The model for a higher primary mass of 12 M$_{\sun}$ was also calculated and are presented in Fig 11. While the EW curves are the same for these two models (as the geometrical parameters are the same), the RV curve of the high-mass star model shows a large deviation from the observations (see Fig. 11) suggesting a low-mass binary system.  
  
It should be noted that the observed ingress and egress times of the eclipse and the relative position of the primary put stringent conditions on the value of $a_{f}$ and in turn the  size of the disk. Large $a_{f}$ will allow a large disk but can also change the beginning of the eclipse earlier and the end of the eclipse later than observed. For the high-mass stars, our model EW curve constrains that the disk can not be larger than 1.1 times the disk size required for the low-mass star model. But the RV curve requires this factor should be about 2.  Hence our model fit to the EW and the RV curves of the shell lines during eclipse, within its assumptions, favours a low-mass star binary for $\epsilon$ Aur. Taking into account of the enhancement of $s$-process elements reported by Sadakane et al. (2010), the F0Ia primary is most likely in post-AGB stage of evolution.

\section{Discussion}

\subsection{Disk structure}
As mentioned earlier, the EW curves of the K {\sevensize \rm I} and other shell lines show jumps at different locations. This indicates that the disk is not homogeneous and fine structures with density jumps are present in the disk. Presence of rings and gaps in the disk plane were suggested earlier by Leadbeater \& Stencel (2010). Leadbeater et al. (2012) have discussed in detail on the jumps observed in the K {\sevensize \rm I} line EW curve in terms of different structures present in the disk. The ingress region of the disk deviates significantly from an uniform-disk model, showing an asymmetry in the disk column density between the ingress and egress (see also Ake 2006, Parthasarathy \& Frueh 1986). The gas column density at the trailing edge found from our model is roughly twice that of the leading edge. Further, the spectral lines are seen in our data (both from VBT and THO) with measurable strength of EW until JD 2456010, when the primary F0Ia star should be far clear from the disk. This indicates that the disk material does not come to an end abruptly after the fourth contact and the disk may have a halo which is significantly larger than its diameter. As noted earlier, the FWHM curve also shows the presence of gaseous matter around the disk after the end of the photometric eclipse. It was noted by Leadbeater et al. (2012) that the EW of the K {\sevensize \rm I} line gradually increased before the eclipse, for nearly the same duration as seen after the eclipse. This shows that with the K {\sevensize \rm I} line we are seeing through the full thickness of the disk atmosphere and the line profile has contributions from all depths within the absorbing region at a given phase of the eclipse. 

The difference in the observed minimum velocity and a delay of 36 days in the time of occurrence in our measurements between the  K {\sevensize \rm I} line and the Fe {\sevensize \rm I} and Cr {\sevensize \rm I} lines indicates that the K {\sevensize \rm I} line originates relatively in the outer regions when compared to the other two lines in the rotating disk. This is also brought out from our models (see Table 6). The time delay was also noticed earlier by Sadakane et al. (2013), though the value found by them is 40 days. The occurrence of Fe {\sevensize \rm I} and Cr {\sevensize \rm I} lines for only a short duration may imply that these elements are present only in a limited region. 

Ake (2006) obtained $FUSE$ far-UV spectrum of $\epsilon$ Aur which show rich emission line spectrum due to low-lying lines and he discussed one possibility of the emission was due to resonance scattering of photons in the disk from an occulting hot source. Griffin \& Stencel (2013) studied the blue and near-UV spectra of $\epsilon$ Aur obtained during the recent eclipse. They have also studied 130 digitized historic spectra from Mount Wilson (dating back to 1930) and from the DAO (dating from 1971). They found precise repetition of disk-related spectral line changes during the three successive eclipses of $\epsilon$ Aur. They conclude that the structure of the disk does not alter appreciably on a time scale of a century. They find that the disk is having an extensive and optically thin outer layer and a flat structure that is tilted near to edge on. Interferometric (CHARA + MIRC) imaging decisively identified the eclipse-causing object to be an opaque disk. The He I 10830 \AA~ line near the mid-eclipses phases clearly indicates that there is hot object at the centre of the disk which was earlier suggested by Parthasarathy \& Lambert (1983d) based on the analysis of UV (IUE) spectra during the 1982 - 84 eclipse.

\subsection {The evolutionary nature of $\epsilon$ Aur system} 

One of the long standing and key unresolved question of the $\epsilon$ Aur system is whether the F0Ia star a low-mass star or a high-mass star. The choice between these two not only shows the evolutionary status of the binary components, but also constrains the nature of the disk. Following arguments were put forward to support these two cases. 

\subsubsection{High-mass model for $\epsilon$ Aur}

The Hipparcos distance to $\epsilon$ Aur of 650 pc is quite uncertain and a reliable distance estimate is very essential to address the evolutionary nature of the system. Distance estimates using other methods like the spectroscopic method, pulsation theory, membership in an OB association, etc., as discussed by Carroll et al. (1991), were also made and they indicate the distance to be upto 1.3 kpc. Guinan et al. (2012) derive a distance of 1.5 kpc based on the interstellar absorption and reddening. The large distance and hence the large absolute luminosity of the F supergiant supports the high-mass star model (Lissauer \& Backman 1984, Guinan et al. 2012). In addition, the semi-regular light variation of the F0Ia star resembles to those found in luminous A-F supergiants (Carroll et al. 1991). The high-mass star advocates that $\epsilon$ Aur has a proto-planetary disk.  Accurate determination of the mass function indicates that a primary of 12M$_{\sun}$ should have a corresponding secondary of mass 11 M$_{\sun}$. The major problem with the high-mass star model is that the secondary is highly under-luminous for its mass of 12M$_{\sun}$. To explain this luminosity problem, Lissauer \& Backman (1984) proposed that the central object in the disk is a close binary  where the total mass is equally shared by the binary components. Such a configuration can be stable if the close binary had a separation of $\le$ 5 AU (Pendleton \& Black 1983). However, as discussed by Hoard et al. (2010), the disk should be 95$\%$ transparent in this case for the minimum possible circumstellar and interstellar reddening of $E(B-V)$ = 0.45 for $\epsilon$ Aur and thus the disk can not eclipse the F0Ia star. Our present study of the K {\sevensize \rm I} line EW and RV curves during recent eclipse shows that the EW curve does not allow the disk size to be very large as required by the RV curve, and the models we worked out do not favour high-mass for    $\epsilon$ Aur.

\subsubsection{Low-mass model for $\epsilon$ Aur: the case of post-AGB object} 

Eggleton \& Pringle (1985) proposed the F0Ia star to be a low-mass post-AGB star to account for the luminosity problem of the secondary. The low-mass star advocates that $\epsilon$ Aur has an accretion disk. Earlier discussion on the low-mass star model given by Takeuti (1986). Analysis of the radial velocities by Lambert \& Sawyer (1986) showed a mass for the secondary between 3M$_{\sun}$ to 6M$_{\sun}$, supporting a low-mass primary. Similar conclusion was also arrived by Hinkle \& Simon (1987) from their accurate two micron CO radial velocity measurements.

The interferometric images of $\epsilon$ Aur during ingress were published by Kloppenborg et al. (2010) which showed that the eclipsing body is an opaque disk which was modelled as a cylinder. The mass ratio of the primary to the secondary obtained by them from binary  solution is 0.62 which constrains a lower-mass for the F0Ia primary of $\epsilon$ Aur. Stencel (2013) summarized the results from the extensive multi-wavelength observations of $\epsilon$ Aur during the recent eclipse.  Hinkle \& Simon (1987) showed from their CO infrared molecular line observations of F0Ia star that the isotopic abundance ratio of $^{12}$C/$^{13}$C = 10 $\pm$ 3 indicating a late evolutionary stage for the primary. The material in the disk seems to be over abundant in $^{13}$C and rare-earth elements which may be the result of accretion and/or mass transfer during the AGB and post-AGB stage of the primary (see Griffin \& Stencel 2012; Stencel 2013). Mourard et al. (2012) from the VEGA/CHARA visibility measurements of $\epsilon$ Aur from 2009 to 2011 eclipse find that the formation of emission wings of H$\alpha$ in an expanding zone almost twice the photospheric size of the F0Ia star. 

Sadakane et al. (2010) obtained from their high resolution spectra taken outside the eclipse the abundances of Mg, Si, S, Ca, Sc, Ti, Cr and Fe and these abundances are close to solar abundances. They found, however, a slight, but definite over-abundances of $s$-process elements Y, Zr and Ba and enhancement of N, Na  in the photosphere of the F0Ia primary. The over-abundances of N, Na and $s$-process elements clearly indicates that F0Ia star has experienced third dredge-up and $s$-process nucleosynthesis. These results support the idea that the F0Ia primary is a post-AGB star. Recently Ishigaki et al. (2012) found similar $s$-process elements over abundance in the photosphere of post-AGB F supergiant CRL 2688. The presence of post-AGB supergiant binaries with dusty disks was first suggested by Parthasarathy \& Pottasch (1986).  High-mass model of F0Ia star cannot explain the over abundance of  $s$-process
elements. For high-mass F0Ia star, by the time it evolves for third dredge up and $s$-process overabundance to occur, carbon burning will takes place and  it will become a Supernova (Iben \& Renzini  1983). During the stable evolution of massive stars third dredgeup and $s$-process nucleosynthesis does not occur (Maeder 1993, Martins 2014).

Muthumariappan \& Parthasarathy (2012) suggested from their 2-D radiative transfer modelling of $\epsilon$ Aur disk that the grains in the disk of $\epsilon$ Aur are most likely amorphous carbon, and it is quite unlikely that silicate dust is present in the disk. This indicates that the dust in the disk was processed and does not show interstellar composition and hence it is not a proto-planetary disk. They concluded that the disk was formed from  mass transfer and/or accretion from the C rich post-AGB star (during slow-wind and/or super-wind mass-loss phase on the AGB, the mass-loss rate estimated by Castelli (1978) from the P-Cygni profiles is $\sim$ 10$^{-7}$ M$_{\sun}$/yr) to its main sequence companion. They have also suggested very extended gaseous envelope around the F0Ia star. Griffin \& Stencel (2013) have discovered that the disk is receiving from the F0Ia primary a very confined stream of material that is enriched in rare-earth elements. This mass-transfer stream is visible between egress phases of the eclipse. Griffin \& Stencel (2013) conclude from their study that the F0Ia primary may be a horizontal branch star. 

The above mentioned recent results on $\epsilon$ Aur system indicate that the F0Ia star is a post-AGB star. Modelling of the EW and the RV curves during the recent eclipse carried out in our study supports that it is a low-mass binary system. We argue that  the disk of $\epsilon$ Aur was formed very recently when the post-AGB primary filled its Roche lobe and transferred mass to the secondary, first suggested by Eggleton \& Pringle (1985). 

\section{Conclusions}

We present high-resolution (30\ 000 \& 70\ 000 from VBT and R = 45\ 000 from TO) and  medium resolution (22\ 000 from THO and 10\ 000 from archive) spectroscopic observations of the eclipsing binary $\epsilon$ Aur taken during 2009-2011 eclipse. Strong variability in the shell spectral lines were seen both in their line profile shape and in their radial velocity. Low-excitation lines of Fe {\sevensize \rm I} and Cr {\sevensize \rm I} appear only for a short duration near the third contact. Non-uniformity in the disk structure is clearly seen from the EW and FWHM variations of the K {\sevensize \rm I} line. It is also observed that the shell lines are present in the disk much after the end of the optical eclipse indicating an extended halo around the disk.  Using a geometrical model, we fit the EW and RV curves of the K {\sevensize \rm I} derived from 346 data points, Fe {\sevensize \rm I} lines and the Na D$_{2}$ line obtained from our data and from the archival data. The model for the K {\sevensize \rm I} line shows that the shell absorption occurs in a transparent atmosphere of diameter 8.9 AU around an opaque disk of diameter about 5.8 AU.  It constrains that the $\epsilon$ Aur binary system has low-mass constituents with the F0Ia primary having a mass of 2.5 M$_{\sun}$ and the mass of the hot star at the centre of the disk is 5.4 M$_{\sun}$. Our results and the overabundance of $s$- process elements found  by Sadakane et al. (2010) constrain that the F0Ia primary is a post-AGB star.

\section {Acknowledgements}

We thank the VBT observing staff for their co-ordination during our observations. We  thank C. Buil, T. Garrel and O. Thizy of the International epsilon Aurigae Campaign 2009 - 2011 for allowing access to their spectra. 
Part of the work on this paper was carried during the MP's visiting professor position in IUCAA. MP is thankful to Prof. Ajit Kembhavi, Prof. Kandaswamy Subramanian and Prof. T. Padmanabhan for their kind encouragement, support and hospitality. We thank the anonymous referee for his/her useful comments.

\newpage
\vspace{15cm}

\begin{table*}
\caption{Table 1. Photemetric contact details of $\epsilon$ Aur during 2009 - 2011 eclipse taken from Stencel (2012)} 
\label{tbl1}
\vspace{1cm}
\begin{tabular}{|c|c|c|}
\hline
Contact   & Date  & Julian Date  \\      
\hline
 I        & 16-08-2009 & 2455070 \\
 II       & 22-02-2010 & 2455250 \\
 III      & 27-02-2011 & 2455620 \\
 IV       & 26-08-2011 & 2455800 \\
\hline  
\end{tabular} 
\end{table*}

\begin{table*}
\begin{minipage}{120mm}
\label{tbl2}
\caption{Table 2. EW and RV measurements of the K {\sevensize \rm I}  line during the recent eclipse of $\epsilon$ Aur  using high-resolution spectra obtained from VBT (see the text for details)} 
\vspace{1cm}
\begin{tabular}{@{}lrrrrrrr}
\hline
\hline
JD & Resol- & EW & RV &JD & Resol- & EW & RV \\
 --245000  & ution            &(\AA)& km s$^{-1}$ &-- 245000  & ution &(\AA)& km s$^{-1}$ \\
\hline
   5556.226 & 30 k &   0.769 &  -39.05 & 5640.100  &75 k&  0.765   & -40.53 \\
   5556.319 & 30 k&   0.759 &   -38.98 & 5640.108  &75 k&  0.786    & -40.74 \\
   5579.270 & 30 k&   0.837 &   -42.04 & 5641.105  &75 k&  0.783   & -40.93 \\
   5579.271& 30 k&   0.847 &   -41.40 &5641.117  &75 k&  0.769   & -40.71 \\
   5581.167 & 30 k&   0.845 &   -40.17 &5642.079  &75 k&  0.775   & -40.27 \\
   5581.172& 30 k&   0.850 &   -39.09 &5642.091  &75 k&  0.784   & -40.49 \\
   5581.194 & 75 k&   0.845 &   -39.44 &5643.081  &75 k&  0.765   & -40.71 \\
   5583.206 & 75 k&   0.777 &   -40.12 & 5643.092  &75 k&  0.767   & -40.63\\
   5583.215 & 75 k&   0.773 &   -40.12 &5643.102  &75 k&  0.786    & -40.47 \\
   5585.219 & 30 k&   0.855&   -39.53 & 5643.113  &75 k&  0.772   & -40.68 \\
   5585.225 & 30 k&   0.856 &   -39.17 & 5645.08  &75 k&  0.759   & -40.97 \\
   5611.264 & 75 k&   0.853  &   -41.52 & 5646.078  &30 k&  0.735   & -37.87 \\
   5612.120 & 75 k&   0.833  &   -40.64 & 5648.074  &30 k&  0.747   & -39.44\\
   5612.126 & 75 k&   0.841 &   -40.87 & 5649.069 &30 k&   0.759  &  -32.25  \\
   5612.132 & 75 k&   0.853  &   -41.31 & 5651.067 &75  k&   0.723 &    -39.75 \\
   5613.238 & 75 k&   0.860 &   -41.16 & 5652.079 &75 k&   0.738&    -39.74 \\
   5613.248 & 75 k&   0.860 &   -41.42 & 5653.067 &75  k&   0.727&    -39.99 \\
   5614.173 & 75 k&   0.853  &   -40.93 &5655.067 &30  k&   0.715&    -38.23 \\
   5614.186 & 75 k&   0.853 &   -41.25 &5678.072 &30  k&   0.653&    -37.10 \\
   5634.085 & 30 k&   0.747 &   -37.40 &5678.073 &30  k&   0.631&    -37.60 \\
   5634.089 & 30 k&    0.764 &    -38.07&5873.394& 75 k &   0.148 &   -28.19 \\
   5635.099 & 30  k&  0.754   & -37.49 &5875.285  & 75 k&   0.128  &   -28.41\\
   5636.172 & 30 k&  0.748   & -39.22 &5881.240 & 75 k&   0.135&   -27.85 \\
   5637.092 & 75 k&  0.784   & -39.78 &5901.279 & 30 k &   0.135 &   -24.93 \\
   5639.198  &75 k&  0.780   & -40.76 &5908.165 & 75 k &   0.112 &   -25.36 \\
   5639.107  &75 k&  0.772   & -40.73 &5909.265 & 75 k &   0.100  &   -26.26 \\
\hline
\end{tabular}
\end{minipage}
\end{table*}

\begin{table*}
\begin{minipage}{120mm}
\caption{Table 3. EW and RV measurements of the K {\sevensize \rm I} line during the recent eclipse of $\epsilon$ Aur  using medium-resolution spectra obtained from Three Hills Observatory } 
\vspace{1cm}
\label{tbl3}
\begin{tabular}{@{}lrrrrrrrr}
\hline
 JD & EW & RV &JD & EW & RV &JD & EW & RV \\
 --245000             &(\AA)& (km s$^{-1}$) &-- 245000 &(\AA)& (km s$^{-1}$)& --245000 &(\AA)& (km s$^{-1}$)\\
\hline
4981	&0.035	&16.4& 5269&	0.453&	19.1& 5510&	0.549	& -28.4 \\
4988	&0.021	&19.1& 5276&	0.447&	19.5& 5516&	0.587	&-30.0 \\
4994	&0.021	&14.8& 5278&	0.460&	18.7& 5520&	0.604	&-30.8\\
5008	&0.032	&14.0& 5288&	0.469&	16.8& 5524&	0.611	&-32.3 \\
5032	&0.072	&17.9& 5290&	0.451&	16.8& 5528&	0.636	&-33.1 \\
5043	&0.077	&19.5& 5294&	0.454&	15.2& 5533&	0.639	&-34.3	\\
5051	&0.089	&18.3& 5299&	0.465&	16.4& 5535&	0.650	&-33.9	\\
5054	&0.080	&19.1& 5301&	0.467&	14.4& 5538&	0.646	&-34.3	\\
5064	&0.096	&18.3& 5303&	0.471&	14.8& 5542&	0.666	&-35.1	\\
5072	&0.128	&18.7& 5307&	0.467&	14.0& 5545&	0.661	&-34.7	\\
5083	&0.162	&18.7& 5313&	0.469&	14.4& 5547&	0.681	&-35.1	\\
5087	&0.169	&18.3& 5324&	0.434&	13.6& 5549&	0.656	&-35.1 \\	
5090	&0.176	&18.7& 5325&	0.425&	11.7& 5551&	0.659	&-35.5 \\	
5105	&0.244	&18.3& 5327&	0.405&	12.9& 5555&	0.672	&-36.2 \\	
5108	&0.252	&17.9& 5329&	0.445&	13.6& 5567&	0.713	&-37.4 \\	
5116	&0.301	&17.9& 5331&	0.432&	12.9&  5568&	0.719	&-37.0 \\	
5127	&0.319	&18.3& 5334&	0.416&	11.3&  5571&	0.725	&-37.0 \\	
5133	&0.329	&18.3& 5350&	0.417&	5.1 &5578&	0.784	&-37.0 \\
5138	&0.327	&19.5& 5359&	0.440&	5.5& 5579&	0.746	&-37.0 \\	
5144	&0.320	&19.5& 5363&	0.400&	3.9& 5580&	0.767	&-37.0 \\	
5150	&0.318	&19.9& 5366&	0.433&	7.4& 5581&	0.773	&-37.4 \\	
5159&	0.315&	19.5 &5369&	0.432&	3.5& 5583&	0.746	&-37.0 \\	
5162&	0.344&	19.5 & 5373&	0.417&	4.7& 5588&	0.752	&-37.8\\	
5163&	0.341&	19.1& 5382&	0.350&	-0.8& 5590&	0.765	&-38.2 \\	
5166&	0.361&	18.3&  5383&	0.394&	2.3& 5591&	0.772	&-38.2\\	
5169&	0.348&	19.9& 5390&	0.386&	2.3& 5594&	0.763	&-38.6\\	
5177&	0.380&	19.5& 5399&	0.368&	-2.3& 5600&	0.782	&-38.6\\	
5185&	0.390&	19.1& 5400&	0.338&	-1.9& 5609&	0.768	&-37.8\\	
5192&	0.387&	19.9& 5405&	0.332&	-5.1& 5615&	0.782	&-38.6\\	
5193&	0.382&	19.5& 5407&	0.356&	-7.4& 5619&	0.796	&-38.6\\	
5194&	0.389&	19.9& 5413&	0.337&	-7.8& 5620&	0.776	&-38.6\\	
5199&	0.395&	19.1& 5414&	0.364&	-4.7& 5621&	0.782	&-38.6\\	
5202&	0.386&	19.5& 5418&	0.321&	-7.8& 5621&	0.750	&-38.2\\	
5205&	0.397&	20.3& 5423&	0.346&	-6.6& 5622&	0.769	&-38.6\\	
5214&	0.395&	19.1& 5424&	0.330&	-10.5& 5622&	0.768	&-37.8\\	
5215&	0.393&	19.5& 5429&	0.360&	-11.3& 5623&	0.761	&-38.2\\	
5224&	0.423&	20.& 5434&	0.362&	-11.7& 5624&	0.758	&-37.8\\	
5228&	0.420&	19.9& 5438&	0.357&	-12.5& 5625&	0.769	&-37.4\\	
5230&	0.420&	19.5& 5440&	0.375&	-12.9& 5626&	0.770	&-38.6\\	
5234&	0.420&	19.9& 5451&	0.425&	-15.2& 5627&	0.755	&-38.6\\	
5245&	0.465&	17.9& 5454&	0.404&	-17.5& 5628&	0.731	&-37.4\\	
5246&	0.469&	18.3& 5455&	0.405&	-19.1& 5629&	0.731	&-37.4\\	
5248&	0.457&	17.9&   5464&	0.397&	-20.3& 5638&	0.712	&-39.0\\	
5250&	0.476&	17.5&   5469&	0.429&	-22.2& 5639&	0.707	&-39.7\\	
5258&	0.445&	18.3&   5477&	0.435&	-24.5& 5641&	0.712	&-37.8\\	
5259&	0.464&	17.9&  5486&	0.442&	-23.4& 5643&	0.704	&-38.2\\	
5261&	0.462&	19.5&  5488&	0.474&	-24.9& 5646&	0.692	&-40.1\\	
5262&	0.466&	19.1&  5493&	0.480&	-25.7& 5646&	0.687	&-38.2\\	
5263&	0.451&	19.1&  5494&	0.479&	-26.5& 5649&	0.677	&-40.5\\	
5266&	0.438&	18.3&  5500&	0.498&	-26.5& 5650&	0.664	&-37.4\\	
\hline
\end{tabular} 
\end{minipage}
\end{table*}

\begin{table*}
\begin{minipage}{120mm}
\label{tbl3}
\caption{Table 3. continued}
\vspace{1cm}
\begin{tabular}{@{}lrrrrrrrr}
\hline
 JD & EW & RV &JD & EW & RV &JD & EW & RV \\
 --245000             &(\AA)& (km s$^{-1}$) &-- 245000 &(\AA)& (km s$^{-1}$)& --245000 &(\AA)& (km s$^{-1}$) \\
\hline
5654&	0.658&	-37.0& 5746&	0.420&	-33.5&	5871&	0.147&	-25.7\\
5655&	0.683&	-40.5& 5750&	0.416&	-33.5&   5872&	0.141&	-26.1\\
5659&	0.660&	-39.0 &5755&	0.426&	-32.0&   5876&	0.148&	-25.3\\
5660&	0.661&	-37.0& 5761&	0.425&	-31.2&   5877&	0.141&	-26.1\\
5662&	0.651&	-39.0 &5765&	0.413&	-31.2&   5879&	0.140&	-25.7\\
5664&	0.644&	-38.6& 5765&	0.410&	-30.8&   5881&	0.134&	-26.1\\
5668&	0.624&	-37.8& 5766&	0.395&	-30.8&  5885&	0.126&	-24.9\\
5669&	0.631&	-37.8& 5767&	0.428&	-31.2&   5888&	0.110&	-24.5\\
5671&	0.631&	-37.8& 5769&	0.397&	-30.4&   5891&	0.116&	-25.3\\
5674&	0.622&	-38.6& 5772&	0.400&	-30.8&   5893&	0.111&	-23.4\\
5675&	0.621&	-38.2& 5772&	0.390&	-30.4&   5897&	0.100&	-23.8\\
5677&	0.601&	-37.8& 5776&	0.383&	-29.6&   5905&	0.090&	-23.0\\
5678&	0.617&	-37.4 &5779&	0.368&	-29.2&   5910&	0.089&	-23.8\\
5680&	0.617&	-37.4 &5782&	0.349&	-30.0&   5912&	0.079&	-23.8\\
5680&	0.599&	-34.7 &5788&	0.315&	-29.6&   5928&	0.054&	-21.4\\
5682&	0.620&	-37.8 &5790&	0.301&	-30.4&   5932&	0.062&	-23.4\\
5684&	0.609&	-37.0 &5791&    0.300&	-30.0&   5936&	0.053&	-23.0\\
5685&	0.615&	-37.0 &5795&	0.289&	-29.6&   5939&	0.054&	-21.0\\
5685&	0.618&	-35.1 &5801&	0.264&	-29.2&   5941&	0.053&	-20.3\\
5686&	0.583&	-35.1 &5809&	0.256&	-28.4&   5948&	0.039&	-21.0\\
5691&	0.596&	-35.5 &5811&	0.249&	-28.8&   5954&	0.038&	-18.7\\
5694&	0.617&	-35.1 &5815&	0.241&	-28.4&   5959&	0.039&	-18.3\\
5700&	0.595&	-35.5 &5819&	0.233&	-28.4&   5962&	0.039&	-20.7\\
5702&	0.542&	-35.1 &5825&	0.231&	-27.7&   5965&	0.038&	-16.8\\
5705&	0.533&	-35.8 &5829&	0.229&	-27.7&   5971&	0.042&	-12.5\\
5714&	0.503&	-35.1 &5832&	0.227&	-26.9&   5974&	0.029&	-14.0\\
5715&	0.486&	-34.3 &5838&	0.210&	-26.5&   5976&	0.031&	-16.8\\
5716&	0.495&	-34.3 &5841&	0.194&	-27.3&   5988&	0.027&	-16.0\\
5724&	0.460&	-34.3 &5842&	0.192&	-26.9&   5989&	0.032&	-17.1\\
5729&	0.450&	-35.1 &5851&	0.157&	-25.7&   6005&	0.021&	-20.3\\
5736&	0.419&	-34.7 &5854&	0.163&	-26.5&   6008&	0.033&	-24.5\\
5740&	0.412&	-34.7 &5860&	0.159&	-26.5&        &      &  \\
5742&	0.423&	-33.1 &5867&	0.163&	-25.7&        &      &  \\
\hline
\end{tabular} 
\end{minipage}
\end{table*}

\begin{table*}
\begin{minipage}{120mm}
\label{tbl4}
\caption{Table 4.  EW and RV measurements of the K {\sevensize \rm I} line during the recent eclipse of $\epsilon$ Aur  using high-resolution spectra obtained by  Potravnov \& Grinin (2013)} 
\vspace{1cm}
\begin{tabular}{@{}lrrrrr}
\hline
 JD & EW & RV &JD & EW & RV \\
 --245000             &(\AA)& (km s$^{-1}$) &-- 245000 &(\AA)& (km s$^{-1}$) \\
\hline
5160.260   &     0.291 &  18.9 &   5538.286   &     0.604 &  -33.6   \\
5163.392   &     0.320  &  19.6 &   5551.343   &     0.608 &  -35.0  \\
5164.336   &     0.324 &  19.9 &   5553.559   &     0.610  &  -35.3  \\
5169.312   &     0.329 &  18.7 &   5554.576   &     0.630  &  -35.3   \\
5176.251   &     0.338 &  19.8 &   5815.366   &    0.210   & -29.2   \\
5188.252   &     0.382 &  20.6 &   5816.467   &     0.215 &  -28.8  \\
5189.195   &     0.365 &  20.2 &   5817.431   &     0.230  &  -29.1  \\
5275.194   &     0.400 &  21.8 &   5818.431   &     0.213 &  -28.7  \\
5276.171   &     0.420  &  21.0 &   5819.492   &     0.216 &  -27.7  \\
5277.296   &     0.432 &  21.6 &   5821.553   &     0.225 &  -27.6 \\
5279.190   &     0.428 &  22.3 &   5823.447   &     0.224 &  -27.2  \\
5431.545   &     0.290  &  -9.0 &   5824.462   &     0.226 &  -27.2  \\ 
5437.504   &     0.345 &  -11.3&   5825.429   &     0.214 &  -28.2  \\
5456.503   &     0.347 &  -17.0&   5827.431   &     0.226 &  -27.6  \\
5459.538   &     0.334 &  -17.7&   5879.290   &     0.114 &  -24.2  \\
5460.595   &     0.347 &  -17.0&   5880.330   &     0.114 &  -24.3  \\
5461.601   &     0.346 &  -16.8&   5881.401   &     0.115 &  -23.7  \\
5463.445   &     0.366 &  -17.7&   5883.403   &     0.110  &  -24.4  \\
5464.591   &     0.384 &  -18.9&   5884.397   &     0.127 &  -25.7  \\
5466.563   &     0.370 &  -18.7&   5885.354   &     0.103 &  -24.5  \\
5467.562   &     0.385 &  -19.5&   5886.329   &     0.099 &  -24.3  \\
5535.330   &     0.586 &  -33.3&   5887.349   &     0.107 &  -24.1  \\
5536.353   &     0.588 &  -32.3&   5888.361   &     0.114 &  -24.9  \\
5537.359   &     0.570  &  -33.5&              &           &         \\
\hline
\end{tabular}   
\end{minipage}
\end{table*}

\begin{table*}
\begin{minipage}{120mm}
\label{tbl4}
\caption{Table 5.  EW and RV measurements of the Fe {\sevensize \rm I} 5110.43 \AA\  line during the recent eclipse of $\epsilon$ Aur  using high-resolution spectra obtained from $\epsilon$ Aur database} 
\vspace{1cm}
\begin{tabular}{@{}lrrrrr}
\hline
 JD & EW & RV  &JD & EW & RV \\
 --245000              &(\AA)& (km s$^{-1}$) &-- 245000 &(\AA)& (km s$^{-1}$) \\
\hline
         &     & & & &   \\
5416.557 &       0.029 &  -15.7 &    5586.237   &     0.076 &  -28.4   \\
5418.584  &      0.034 &  -12.4 &    5587.279   &     0.07  &  -28.5  \\
5420.548  &      0.036 &  -14.6 &    5597.284   &     0.063 &  -30.6 \\
5434.563  &      0.04  &  -19.5 &    5599.246   &     0.064 &  -32.5   \\
5439.574  &      0.046 &  -15.0 &    5601.264   &     0.068 &  -32.9 \\
5441.545  &      0.041 &  -15.6 &    5603.266   &     0.073 &  -33.9 \\
5442.556  &      0.045 &  -14.3 &    5604.266   &     0.07  & -32.4  \\
5443.557  &      0.041 &  -14.1 &    5611.256   &     0.079 &  -32.5 \\
5444.595  &      0.036 &  -19.1 &    5613.335   &     0.079 &  -32.3 \\
5450.542  &      0.047 &  -15.6 &   5629.349   &     0.08  &  -32.4 \\
5453.486  &      0.042 &  -13.7 &    5630.427   &     0.08  & -42.5 \\
5454.494  &      0.037 &  -13.6 &    5631.304   &     0.074 &   -40.4\\
5459.487  &      0.04  &  -17.2 &    5642.278    &    0.072 &  -38.7 \\
5460.504  &      0.043 &  -16.5 &    5653.426    &    0.068 &  -44.1 \\
5465.487  &      0.036 &  -18.4 &   5654.302    &    0.064 & -43.2  \\
5466.565  &      0.033 &  -17.7 &    5655.282    &    0.069&   -38.5 \\
5468.453  &      0.035 &  -17.4 &    5656.354    &    0.071 &  -31.8 \\
5469.448  &      0.037 &  -20.8 &    5669.383    &    0.083 &  -45.1 \\
5484.385  &      0.04  &  -29.0 &    5682.349    &    0.076 &  -36.0 \\
5485.476  &      0.047 &  -26.8 &   5697.343    &    0.085 &  -42.9 \\
5488.421  &      0.043 &  -24.5 &    5700.325    &    0.08  &  -43.1 \\
5490.502  &      0.048 &  -27.9 &   5703.331    &    0.079 &  -37.6 \\
5491.426  &      0.041 &  -27.3 &   5704.326    &    0.081 &  -37.1 \\
5495.398  &      0.047 &  -25.7 &   5707.334    &    0.072 &  -33.4 \\
5496.455  &      0.046 &  -23.0 &   5773.544    &    0.037 &  -25.0 \\
5498.404  &      0.051 &  -23.7 &   5778.592    &    0.039 &  -21.3 \\
5508.503  &      0.061 &  -25.2 &   5779.547    &    0.04  &  -18.2 \\
5517.324  &      0.049 &  -21.7 &   5790.515    &    0.029 &  -12.5 \\
5519.503  &      0.048 &  -22.6 &   5801.458    &    0.023 &  -15.8 \\
5520.504  &      0.054 &  -21.9 &   5801.623    &    0.017 &  -21.5 \\
5524.289  &      0.053 &  -23.0 &   5813.507    &    0.02  &  -27.2 \\
5525.300  &      0.052 &  -23.7 &   5817.539    &    0.025 &  -32.0 \\
5527.292  &      0.056 &  -23.0 &   5821.444   &     0.028&   -21.0 \\
5531.920  &      0.045 &  -23.8 &   5828.537    &    0.031 &  -24.2 \\
5535.286  &      0.053 &  -27.3 &   5830.496    &    0.026 &  -28.6 \\
5543.394  &      0.042 &  -30.0 &   5840.477    &    0.034 &  -30.3 \\
5544.259  &      0.039 &  -27.7 &   5844.415    &    0.024 &  -15.6 \\
5545.239  &      0.045 &  -30.1 &   5856.448    &    0.021 &  -21.7 \\
5549.224  &      0.044 &  -31.1 &   5882.307    &    0.026 &  -17.6 \\
5550.224  &      0.048 &  -33.6 &   5889.367    &    0.019 &  -21.6 \\
5551.266  &      0.047 &  -30.5 &   5898.263    &    0.021 &  -23.6 \\
5556.356  &      0.051 &  -34.2 &   5906.388    &    0.014 &  -17.0 \\
5571.254  &      0.069 &  -31.2 &   5918.344    &    0.015 &  -24.2\\
5575.289  &      0.065 &  -29.9 &   5927.258    &    0.026 &  -13.7 \\
5576.249  &      0.072 &  -31.9 &   5938.307    &    0.019 &  -14.1\\
5577.230  &      0.073 &  -31.1 &   5949.268    &    0.025 &  -16.6 \\
5579.317  &      0.073 &  -28.8 &   5957.268    &    0.011 &  -25.7 \\
5581.410  &      0.068 &  -28.2 &   5965.401    &    0.018 &  -19.6 \\
5582.313  &      0.07  &  -30.1 &   5979.439    &    0.02  &  -27.7 \\
5583.274  &      0.07  &  -28.4 &   5993.451    &    0.018 &  -17.5 \\
5585.282  &      0.068 &  -29.0 &  &&\\
\hline
\end{tabular}   
\end{minipage}
\end{table*}

\newpage
\begin{table}
\label{tbl5}
\caption{Table 6. Model parameters of the disk of $\epsilon$ Aur and the binary system}
\vspace{1cm}
\begin{tabular}{|c|c|c|c|}
\hline
Parameter   & Model for  & Model for &Model for  \\      
            &   K I 7699 \AA\ line & Fe I 5110 \AA\ line & Na D$_{2}$ line    \\  
\hline
\noindent
$Adapted:$  &               &        &    \\
            &               &        &    \\
Mass function (M$_{\sun}$) &  2.51      & 2.51  & 2.51     \\
Period  (days)&  9896      &  9896    & 9896   \\
Eccentricity  &  0.227     &  0.227   & 0.227  \\
              &           &        &           \\
$Model$ $derived:$ &      &        &         \\
              &           &        &         \\              
Disk outer Diameter (AU) & 8.9  &  8.7    &  8.1 \\
Disk inner diameter (AU) &  5.8  & 5.4      & 4.1 \\
RV Offset  (km s$^{-1}$) & --4.0     & --3.0     & --1.0 \\
Secondary mass (M$_{\sun}$)& 5.4  & 5.4 & 5.4 \\
Primary mass (M$_{\sun}$)& 2.5  & 2.5 & 2.5 \\
Density function    &  accretion disk  & accretion disk   &  accretion disk     \\
                    &  midplane & midplane & midplane   \\                
\hline  
\end{tabular} 
\end{table}

\vspace{2.0cm} \protect \begin{figure*}
\vspace{10.0cm}
\includegraphics{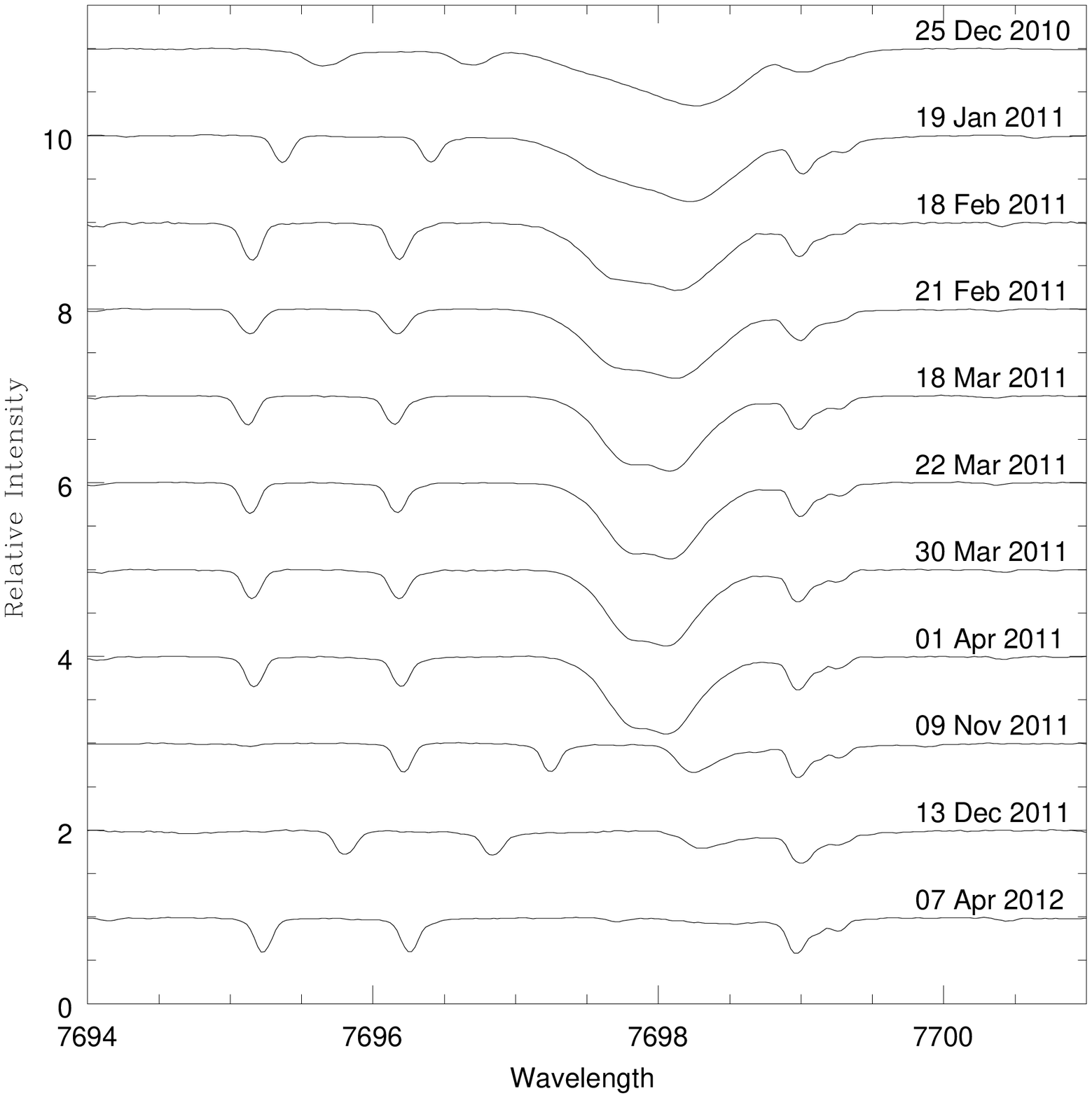}
\vspace{1.0cm}
\caption{A sample of time series spectra from VBT showing the variation of K {\sevensize \rm I} line profile during the eclipse. } 
\end{figure*}

\vspace{2.0cm} \protect \begin{figure}
\vspace{10.0cm}
\includegraphics{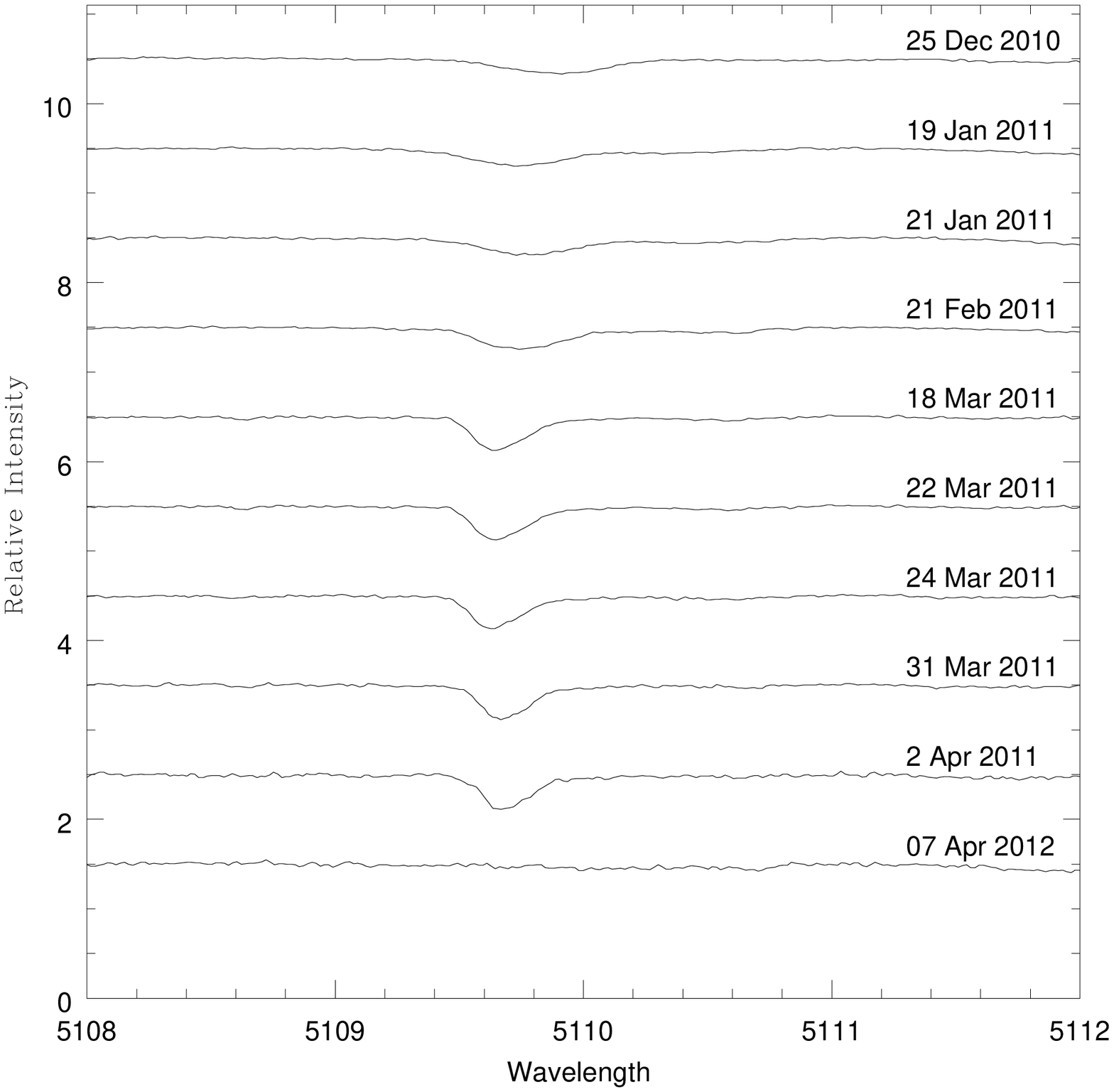}
\vspace{1.0cm}
\caption{A sample of time series spectra from VBT showing the variation of Fe {\sevensize \rm I} 5110.43 \AA\ line profile during the eclipse. } 
\end{figure}
 
\vspace{2.0cm} \protect \begin{figure}
\vspace{10.0cm}
\includegraphics{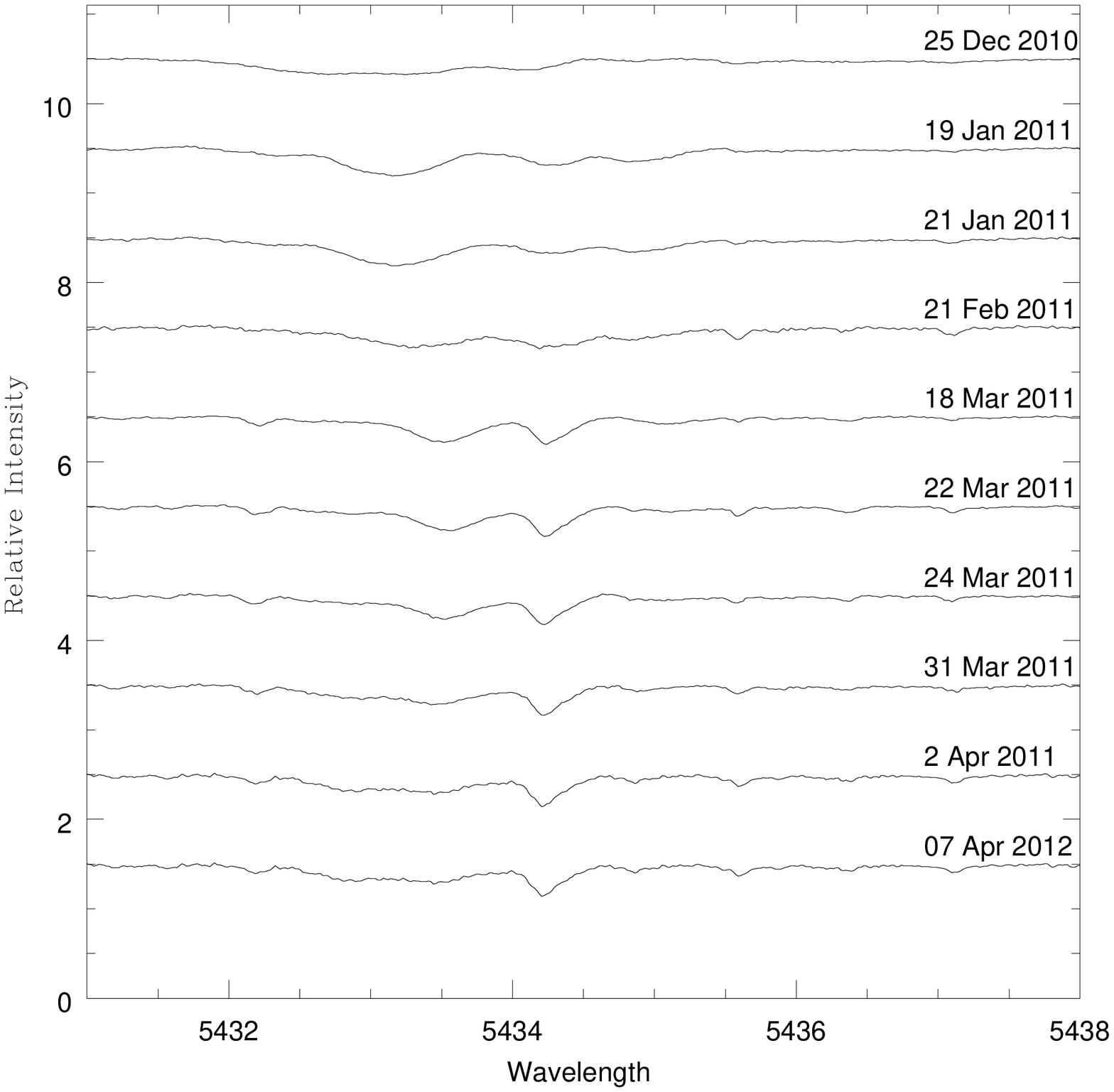}
\vspace{1.0cm}
\caption{A sample of time series spectra from VBT showing the variation of Fe {\sevensize \rm I} 5434.53 \AA\ line profile during the eclipse. } 
\end{figure}

\vspace{2.0cm} \protect \begin{figure}
\vspace{10.0cm}
\includegraphics{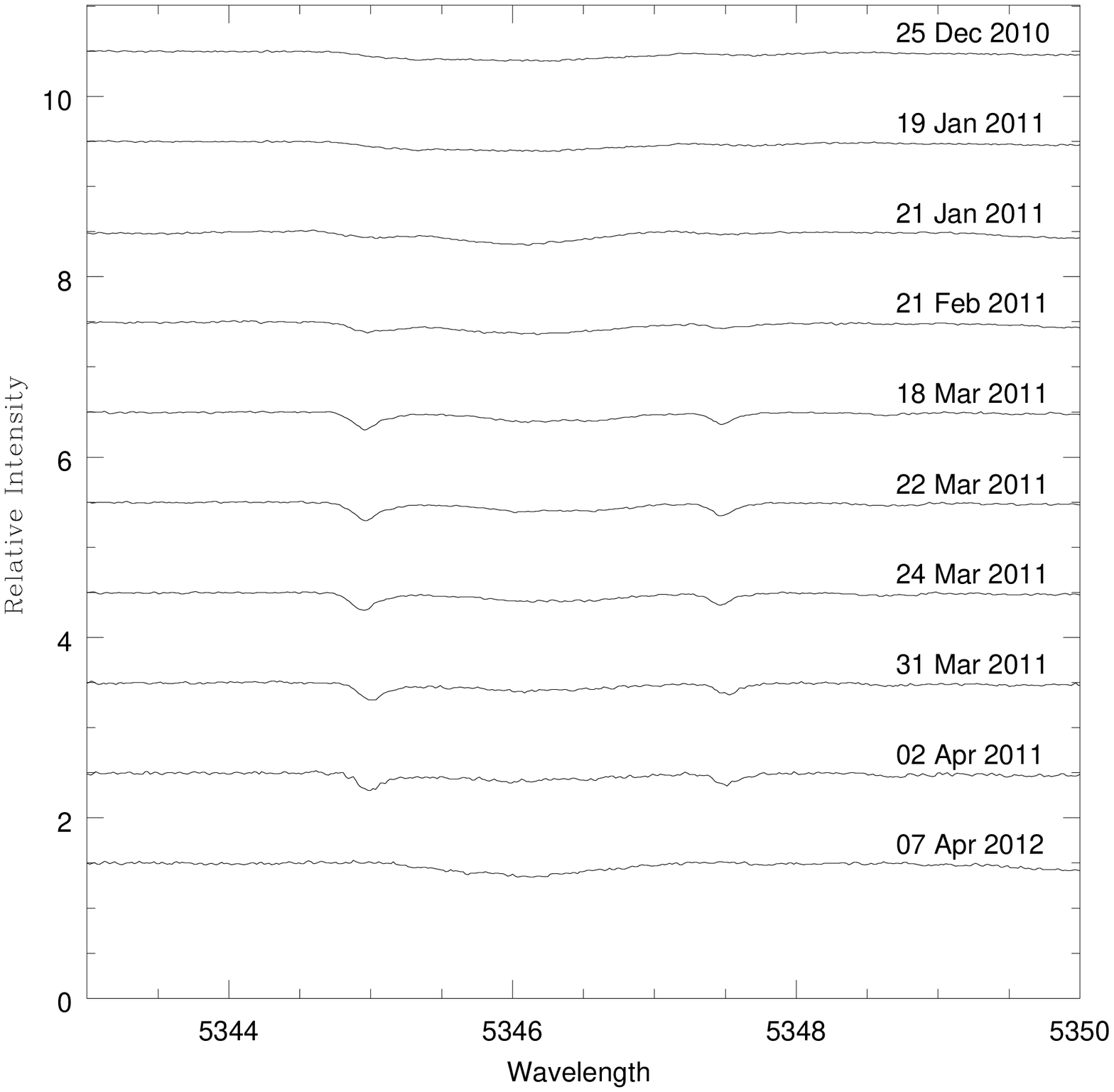}
\vspace{1.0cm}
\caption{A sample of time series spectra from VBT showing the variation of Cr {\sevensize \rm I} line profile during the eclipse. } 
\end{figure}

\vspace{2.0cm} \protect \begin{figure*}
\vspace{10.0cm}
\includegraphics{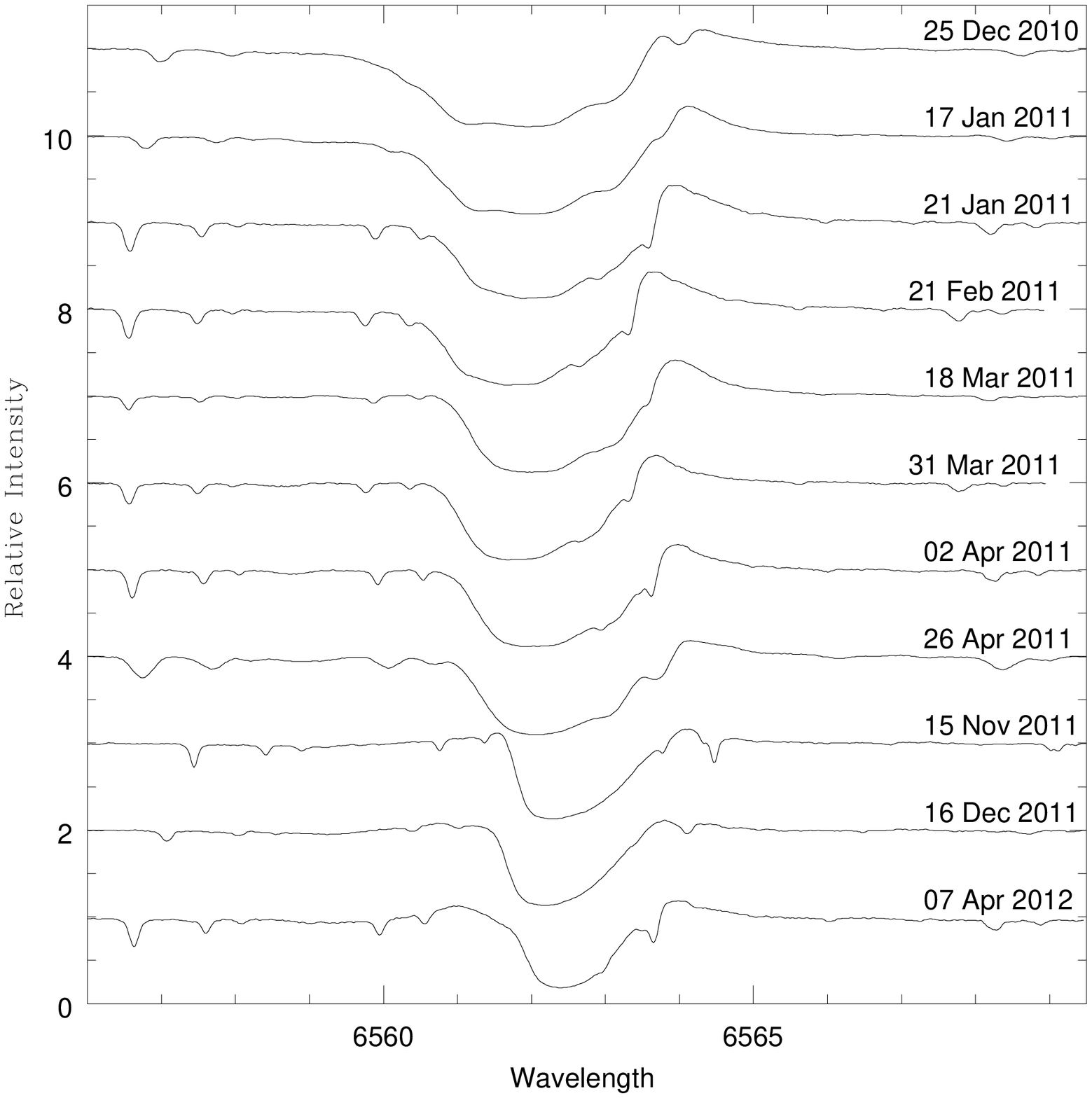}
\vspace{1.0cm}
\caption{A sample of time series spectra from VBT showing the variation of H$\alpha$ line prifile during the eclipse. } 
\end{figure*}

\newpage
\vspace{2.0cm} \protect \begin{figure}
\vspace{10.0cm}
\includegraphics{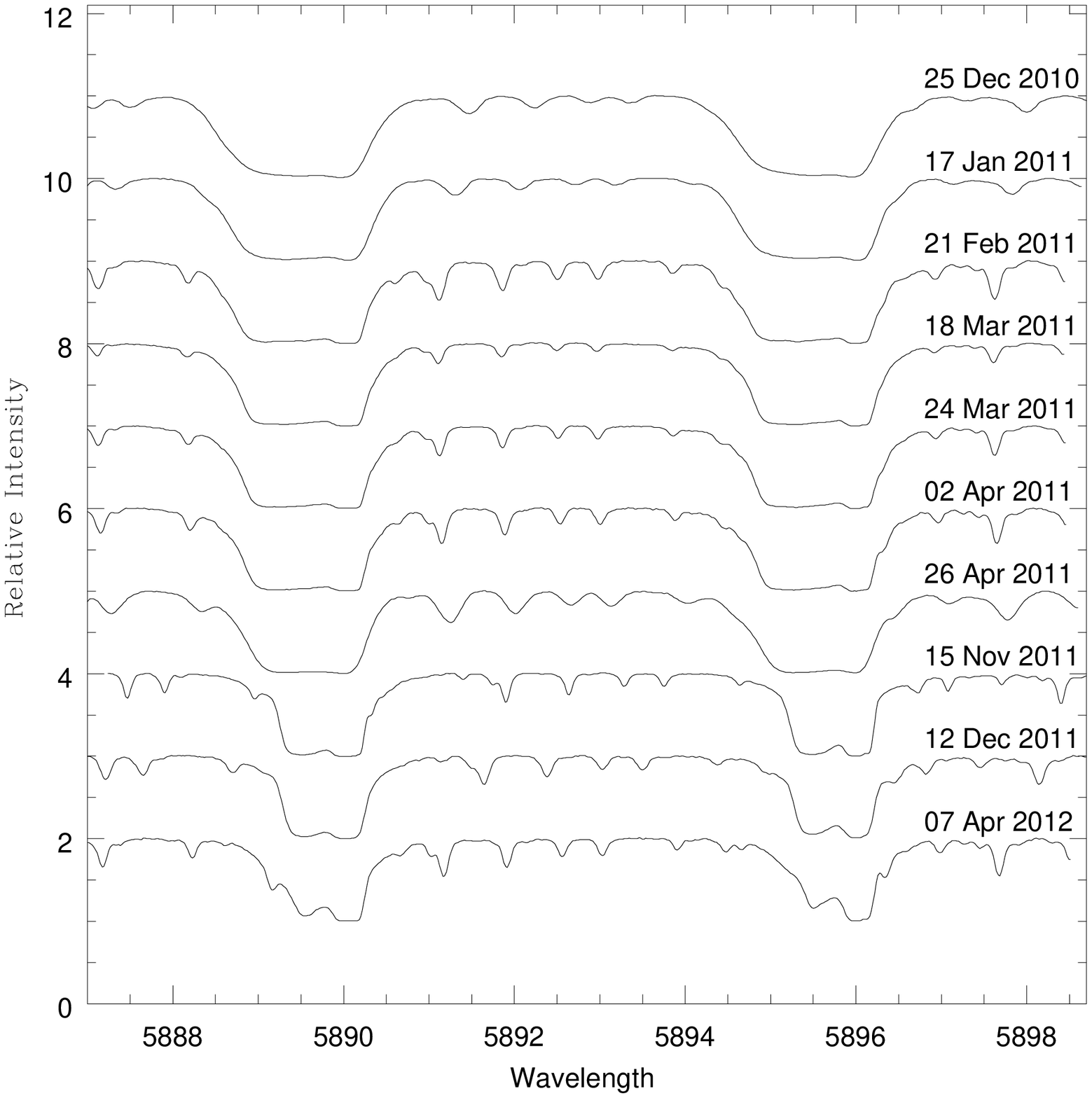}
\vspace{5.0cm}
\caption{A sample of time series spectra from VBT showing the variation of Na D line profiles during the eclipse. } 
\end{figure}
\newpage

\vspace{2.0cm} \protect \begin{figure}
\vspace{10.0cm}
\includegraphics{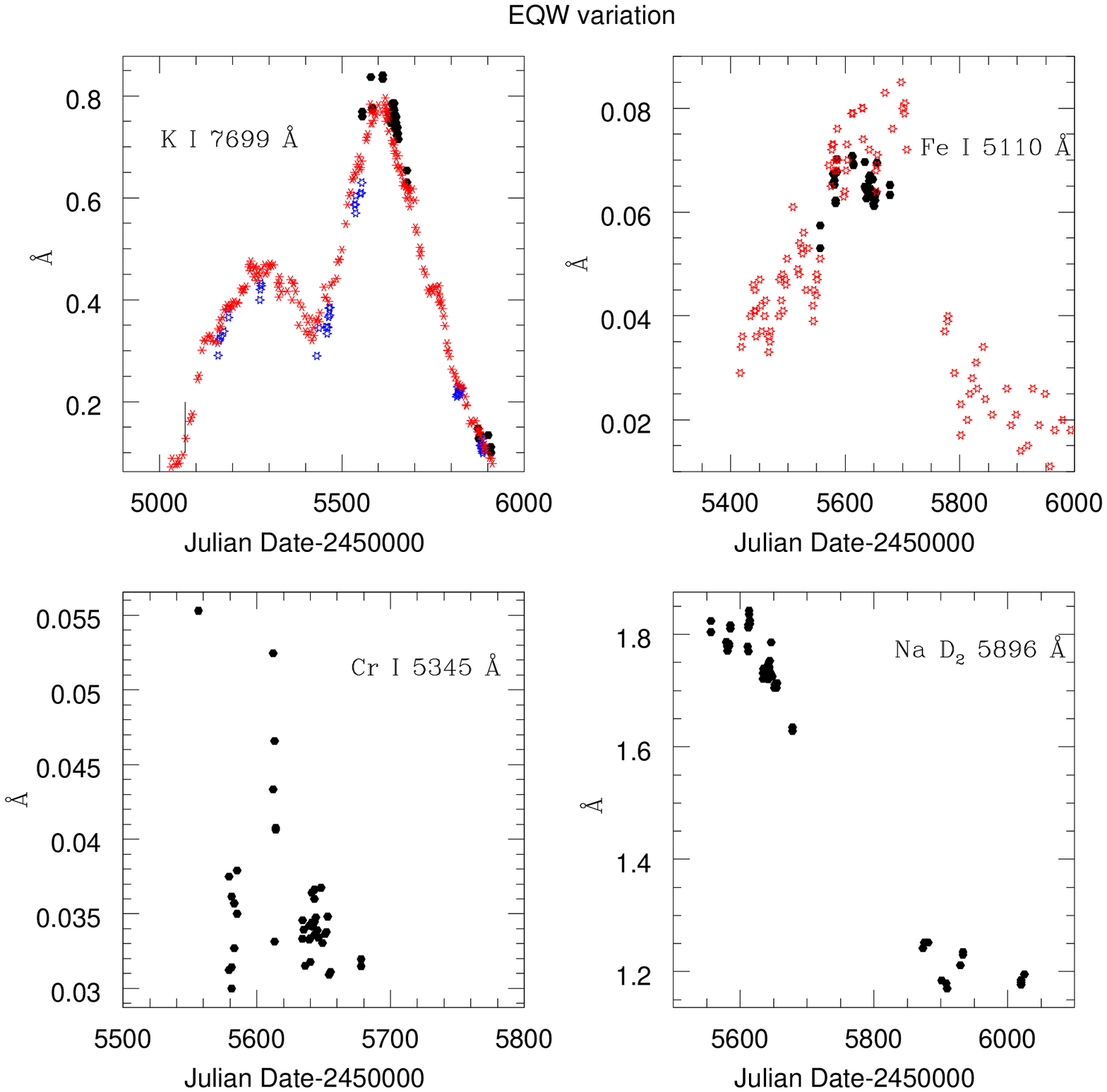}
\vspace{5.0cm}
\caption{Equivalent width variation of the spectral lines during the eclipse. Data description for the K {\sevensize \rm I} line: filled black circles are from VBT observations, red asterisk are from  THO, open blue stars are from TO and open red stars are from the archive.} 
\end{figure}

\vspace{2.0cm} \protect \begin{figure}
\vspace{10.0cm}
\includegraphics{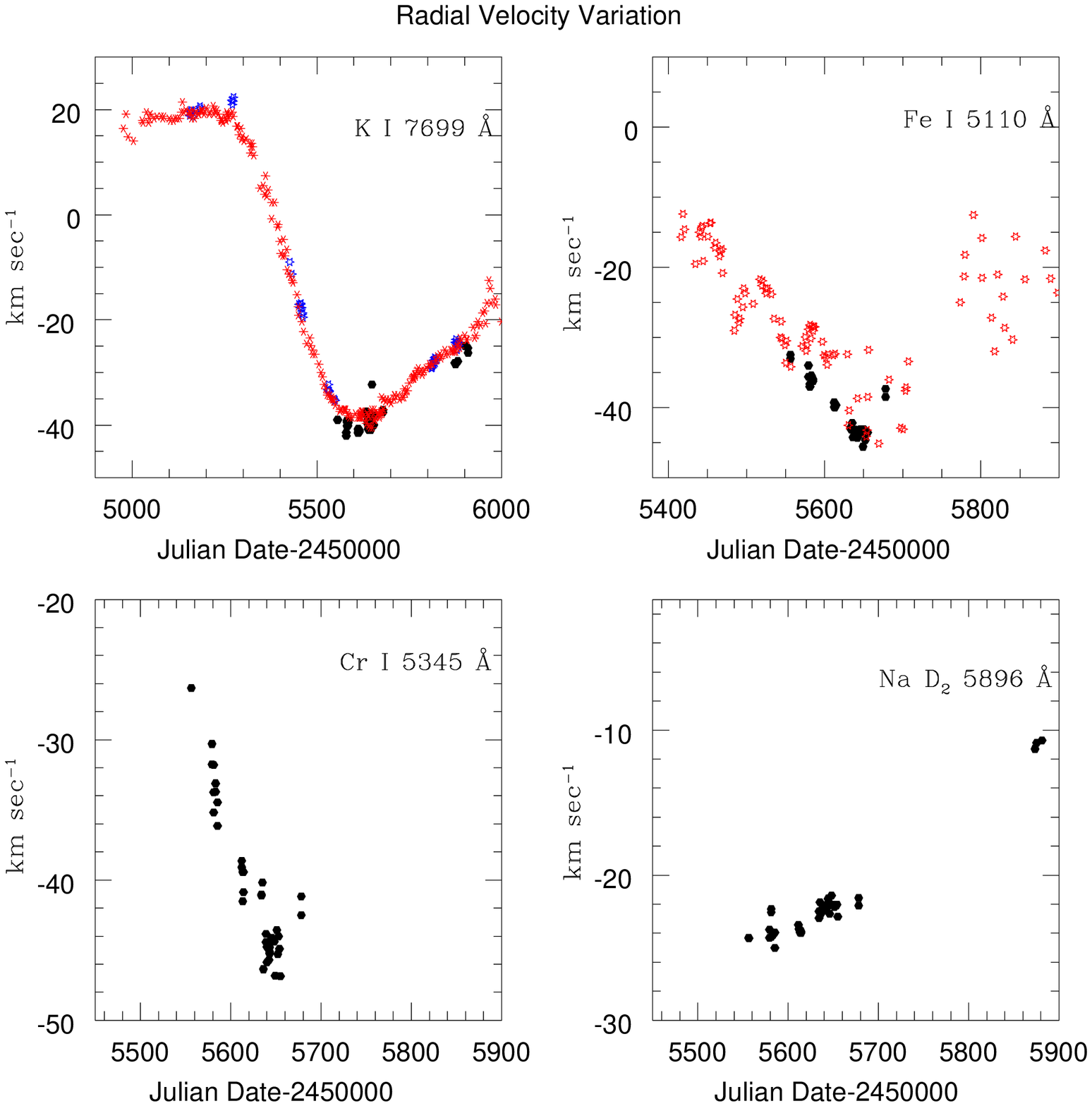}
\vspace{5.0cm}
\caption{Radial velocity variation of the spectral lines during the eclipse. Data description for the K {\sevensize \rm I} line: filled black circles are from VBT observations, red asterisk are from THO, open blue stars are from TO and open red stars are from the archive.} 
\end{figure}

\vspace{2.0cm} \protect \begin{figure}
\vspace{10.0cm}
\includegraphics{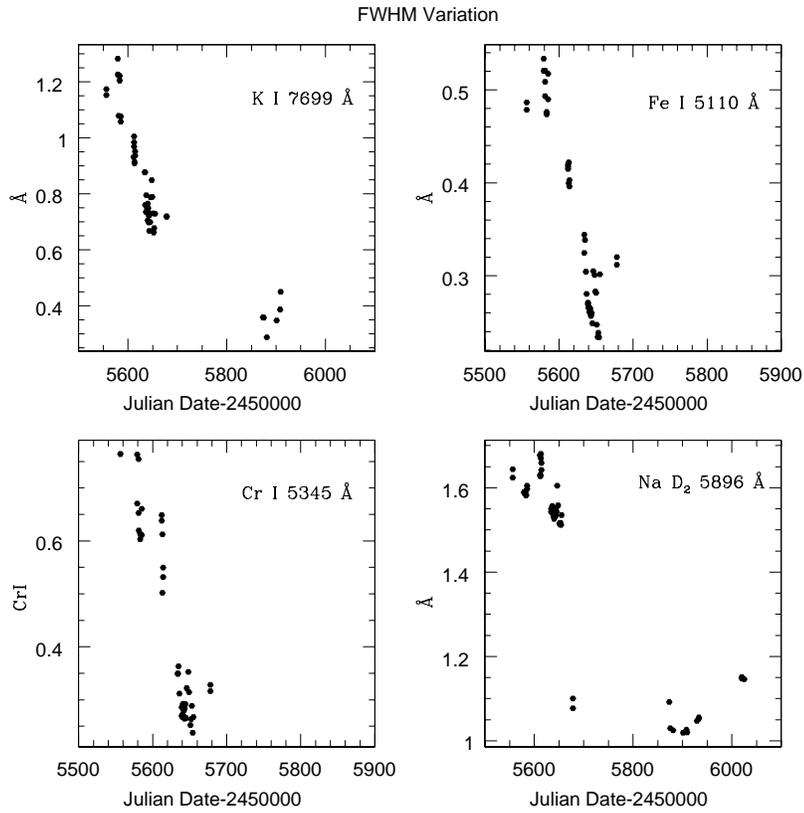}
\vspace{5.0cm}
\caption{Full-width at half-maximum variation of the spectral lines during the eclipse (VBT data).} 
\end{figure}

\vspace{2.0cm} \protect \begin{figure}
\vspace{10.0cm}
\includegraphics{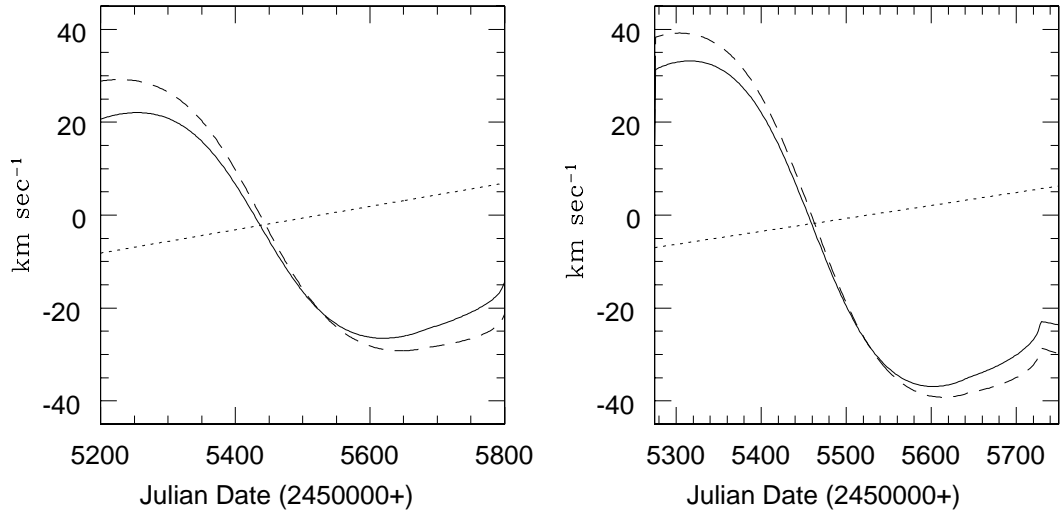}
\vspace{2.0cm}
\caption{Variation of $v_{r,disk}$ (dashed line), $v_{r,sec}$ (dotted line) and their addition (solid line) during the eclipse for the low-mass star model (2.5 M$_{\sun}$, left panel) and for the high-mass star model (12M$_{\sun}$, right panel). See text for details.}
\end{figure}

\vspace{2.0cm} \protect \begin{figure}
\vspace{10.0cm}
\includegraphics{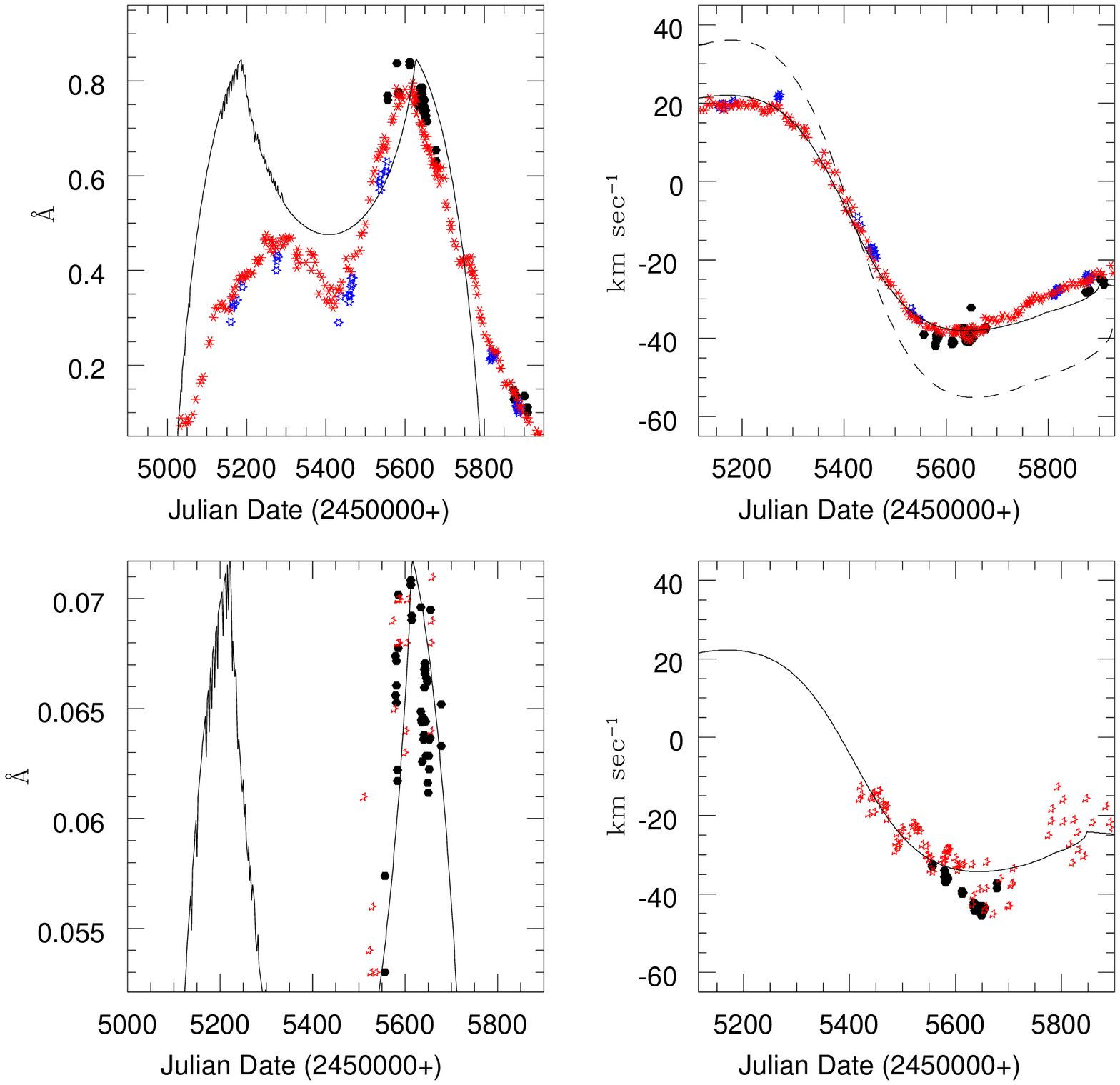} 
\includegraphics{angle=0}
\includegraphics{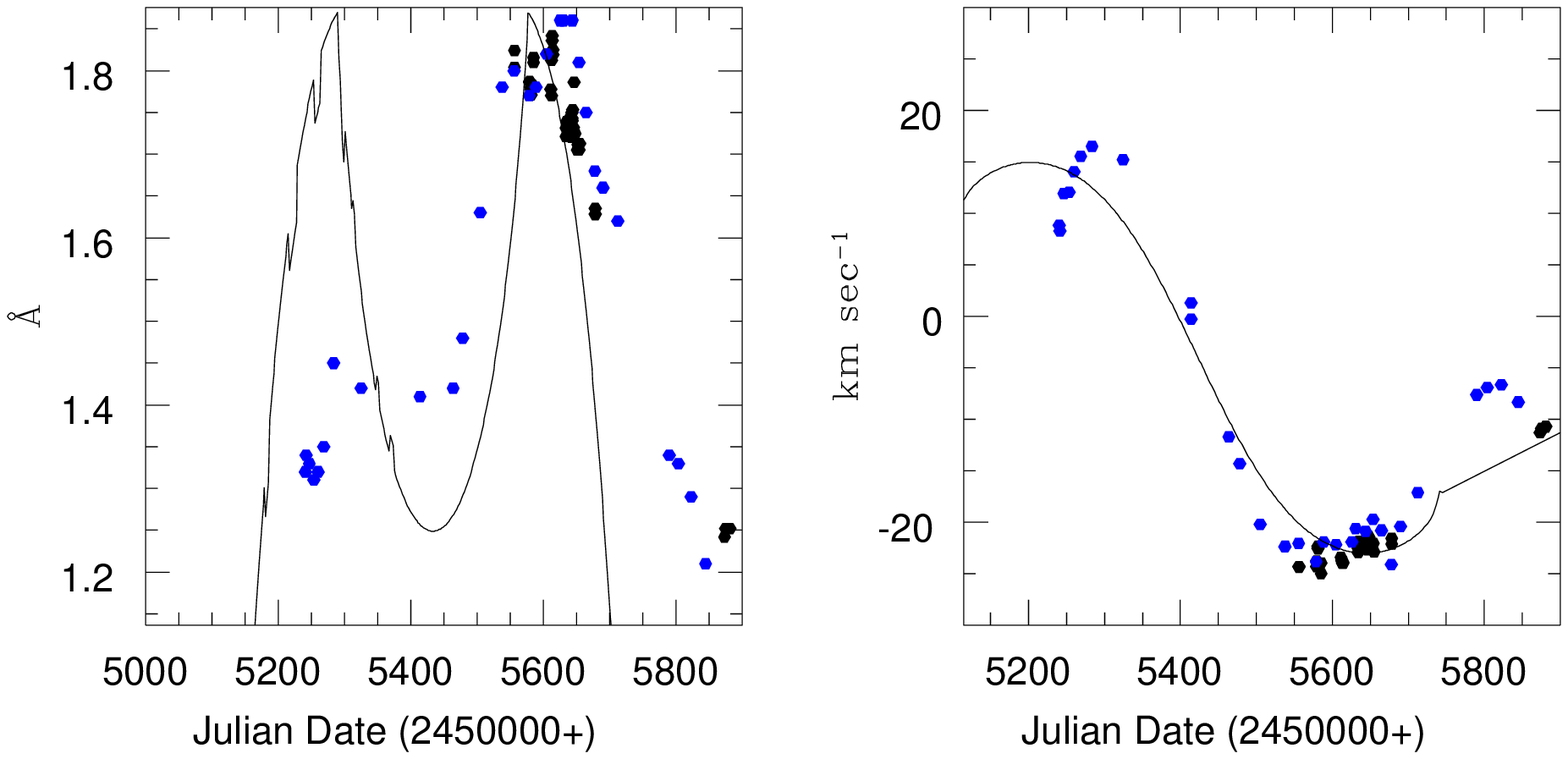}
\vspace{7.0cm}
\caption{A geometrical model fit to the $\epsilon$ Aur data. a) Top row: EW (left) and RV (right) fit to the K {\sevensize \rm I} line; filled black circles are from VBT observations, red asterisk are from THO and open blue stars are from TO. Solid line is our proposed model fit to the data for the primary mass of 2.5M$_{\sun}$, dashed lines is for the primary of mass 12M$_{\sun}$. b) Middle row: EW (left) and RV (right) fit to the Fe I 5110.43 \AA\  line; filled black circles are from VBT observations open red stars are from archive. Solid line is our proposed model fit to the data for the F0Ia mass of 2.5M$_{\sun}$. c) Bottom row: EW (left) and RV (right) fit to the Fe I 5110.43 \AA\ line; filled black circles are from VBT data filled blue hexagons are from Gorodenski 2012. Solid line is our proposed model fit to the data for the F0Ia mass of 2.5 M$_{\sun}$.} 
\end{figure}

\label{lastpage}
\end{document}